\documentclass[letterpaper]{article} 
\usepackage{aaai2027} 
\usepackage{graphicx} 
\usepackage{times,helvet,courier}
\usepackage{microtype}
\usepackage[hyphens]{url}
\usepackage[strings]{underscore}
\usepackage{natbib}
\usepackage{caption}
\usepackage[utf8]{inputenc}
\usepackage{textcomp,textgreek}
\usepackage{graphicx,booktabs,array,tabularx}
\usepackage{amssymb,amsmath}
\usepackage{subfig,xspace}
\usepackage{booktabs}
\graphicspath{{figs/}}
\newcolumntype{L}{>{\raggedright\arraybackslash}X}
\newcommand{\platform}{Telegram\xspace}
\newcommand{\mbfc}{MBFC}
\xspaceaddexceptions{\textquoteright}
\title{The Tag is the Signal: URL-Agnostic Credibility Scoring for Messages on Telegram}

\author{
Yipeng Wang\textsuperscript{\rm 1}, 
Huy Gia Han Vu\textsuperscript{\rm 2}\thanks{Work done at Northeastern University.},
Mohit Singhal\textsuperscript{\rm 1}}

\affiliations{
\textsuperscript{\rm 1}Northeastern University\\
\textsuperscript{\rm 2}Rutgers University-Newark\\

\{wang.yipen,m.singhal\}@northeastern.edu, han.vu@rutgers.edu
}

\begin{document}
\maketitle

\begin{abstract}
Telegram has become one of the leading platforms for disseminating misinformation. However, many existing pipelines still judge message credibility based on domain names or lexical features. Such methods work well on traditional news articles, but high-risk posts on Telegram rarely contain URLs, leading to failures for link-based and standard TF-IDF models. To this end, we propose \textsc{Tag2Cred}, an interpretable representation that factorizes each message into theme, claim framing, call to action, and evidence, assigned via a fine-tuned LLM and mapped to risk scores via logistic regression. This representation supports message-level triage and aggregate monitoring. On the domain-disjoint evaluation of 65,933 messages, \textsc{Tag2Cred} achieves AUC~0.864 and Macro-F1~0.821, outperforming TF-IDF and SBERT. On a native zero-URL benchmark of 1,252 messages, it achieves AUC~0.979 and Macro-F1~0.912. Ablations show that Call to action (CTA) is the strongest single field, but the main performance boost comes from claim framings and cross-field structure. In the monitoring analysis, finance/crypto promotion combined with scarcity narratives dominates the high-risk tail, while public health content shows disproportionate enrichment. The weighted ensemble (TF-IDF + TAG2CRED + SBERT) further improves performance (AUC~0.893; Macro-F1~0.828), indicating that rhetorical tags and lexical features reflect complementary aspects of information risk.
\end{abstract}
\begin{center}
\large
\textcolor{red}{\textbf{This paper has been accepted at ICWSM 2027, please cite accordingly.}
}\end{center}

\section{Introduction}

Social media platforms have become the core infrastructure for information exchange and mobilization~\cite{dechoudhury2016racial,venancio2024telegram,pera2025collective,singhal2023sok,verma2011situational}.
The dissemination of low-credibility information often depends not only on the factuality of the claim itself, but also on the use of rhetorical devices such as assertive language, emotional appeal, selective evidence, and a direct call to action~\cite{da-san-martino-etal-2019-fine,dimitrov-etal-2021-detecting,solovev2022moralemotions,da-san-martino-etal-2020-prta,singhal2023cybersecurity}. For analysts and platform operators, this poses an operational prioritization problem: amid the continuous influx of heterogeneous posts, what information should be prioritized for the in-depth verification and intervention process? Most of the existing studies solve this problem by taking the information source as a proxy indicator of credibility, using media-level resources like Media Bias/Fact Check (MBFC)~\cite{mbfc} and transmitting the source evaluation results to specific posts through URLs and domain names~\cite{baly2018predicting,bozarth2020higherground,weld2021political,volkova2017separating,kumarswamy2025parler,nakov2024survey,samory2020newsphere}.

Telegram messages often lack standard URLs or contain obfuscated links pointing to unrated domains~\cite{benevenuto2024misinfo,roy2025darkgram}.
On the other hand, the Telegram ecosystem is characterized by a long-tail distribution: a few core, large-scale channels account for most of the link traffic, while long-tail channels often spread screenshots, forwarded claims, and unrated domain names~\cite{kireev2025propaganda,steffen2025memes,vafa2025learning}. This leads to overestimating the risk level of low-risk core channels and systematically underestimating the high-risk content in long-tail channels by only evaluating the content through domain-name credibility~\cite{xu2019pumpdump,roy2025darkgram}.

We propose \textsc{Tag2Cred}, a first step toward URL-agnostic credibility-risk scoring on Telegram. Rather than relying on lexical features or source identity, \textsc{Tag2Cred} represents each message as a compact rhetorical tag profile to generate interpretable risk scores for prioritization and triage.
We define a simplified rhetorical tag system through qualitative open coding that covers four dimensions: \textit{theme} (such as Finance/Crypto, Public health \& medicine), \textit{rhetorical stance/claim types} (such as Speculative forecast/prediction, emotional appeal), \textit{call to action} (such as join/subscribe, Buy/invest/donate), and \textit{evidence presentation} (such as external links, charts/screenshots, or no evidence). We collected 9,371,099 Telegram messages from 234 public channels. Our supervised evaluation begins with the resulting MBFC-linkable subset of 87,936 messages (out of the 9.37M collected). After thresholding, 65,933 high-confidence labeled messages remain for supervised training and evaluation. Using an annotated seed Telegram message set, we fine-tuned a large language model (Qwen-32B) as a constrained label-assignment model to output the above labels for new messages. The resulting low-dimensional tag vector explicitly excludes URLs, domain names, and other source-identity features. We trained an $\ell_2$-regularized logistic regression model on the tag vectors. For the grouped evaluation, its predicted probability $\hat{p}(y=1 \mid m) \in [0,1]$ serves as the credibility-risk score.
Our analysis also yields insights into the spread of misinformation on Telegram: \textit{Call to action} (CTA) is the strongest predictor. Moreover, pairing CTA with claim framing and selected grouped cross-field structure further explains the risky signal. Theme and evidence contribute comparatively little under domain-disjoint evaluation but are essential for human monitoring.

Under URL masking and 50 repeated grouped splits, we observed that \textsc{Tag2Cred} raised macro-F1 from 0.761 to 0.821 and lowered the Brier score from 0.174 to 0.153 relative to the URL-masked TF--IDF baseline. A weighted ensemble of TF--IDF, \textsc{Tag2Cred} and SBERT, in the same setting, further achieved AUC 0.893, macro-F1 0.828, and ECE 0.113, indicating that rhetorical tags capture complementary information beyond lexical features.

This paper makes the following contributions:

\begin{enumerate}

\item \textbf{URL-independent rhetorical tagging system for Telegram messages}: built a compact set of interpretable rhetorical tags that covered topics, rhetorical stances, calls to action, and evidence presentation.

\item \textbf{Compact message-profile representation for monitoring}: uses a closed-vocabulary profile over theme, claim framing, call to action, and evidence/support, with both per-field bucket indicators and selected grouped cross-field features, so that the same representation can support calibrated message-level risk triage and aggregate monitoring of recurring strategy types.

\item \textbf{Systematic comparison with benchmark models}: compared \textsc{Tag2Cred} to TF--IDF, SBERT, and an LLM classifier; under domain-disjoint URL masking, \textsc{Tag2Cred} improves macro-F1 by 0.060 over TF--IDF and 0.065 over SBERT; the ensemble attains macro-F1 = 0.828 (ECE = 0.113).

\item \textbf{Telegram strategy profiling and monitoring}: uses the message profile to characterize recurring high-risk strategy prototypes, analyze enrichment patterns, and track temporal drift and bursty shifts in rhetorical packaging.

\end{enumerate}

\section{Related Work}

\textbf{Credibility evaluation based on domain names and links}: Prior works have used various textual and structural features to predict the factual and political bias of information~\cite{baly2018predicting,nakov2024survey,stefanov2020predicting}.
The large-scale analysis of the behavior of the news sharing of the online community also relies on MBFC-style domain tagging to represent bias and factuality~\cite{weld2021political,gruppi2021nela,kumarswamy2025parler,etta2022comparing,patricia2019link}, while the comparative study of real tag lists points out that the selection of reputation resources will affect the results of downstream misinformation measurement~\cite{bozarth2020higherground,bovet2019influence,cinelli2020covid,cinelli2021echo}.

\textbf{Telegram misinformation, short message nature, and URL scarcity}: Scholars have extensively studied Telegram as a carrier of propaganda, conspiracy narratives, and economically motivated manipulation~\cite{benevenuto2024misinfo,kireev2025propaganda,steffen2025memes,vanetik2023propaganda,bareikyte2025digitally}.
Empirical research also shows that Telegram messages are usually short in duration and their structure is significantly affected by channel functionality and forwarding mechanisms, which weakens the effectiveness of URL-based measurement methods and increases the transmission proportion of unrated sources~\cite{benevenuto2024misinfo,kireev2025propaganda,steffen2025memes}. Existing studies have also found widespread link obfuscation, nesting, and cross‑platform collaboration, such as pump‑and‑dump manipulation, darknet markets, malicious file sharing, or fraud ecosystems~\cite{xu2019pumpdump,roy2025darkgram,vafa2025learning}.

\textbf{Style-based credibility and fake news detection}: A large number of studies have explored the possibility of inferring credibility only through writing style~\cite{horne2017justin,rashkin2017truth,volkova2017separating,wang-2017-liar,ott-etal-2011-finding,feng-etal-2012-syntactic}.
Previous research shows that stylistic features can effectively distinguish between extreme partisan media and mainstream media~\cite{potthast2018hyperpartisan,huang2020conquering,hansen-etal-2021-automatic}, but reliability is significantly reduced in the classification task of real news and fake news.
Subsequent studies have expanded this direction through a larger corpus (credibility list annotation based on the source level), using regularized regression and neural models to learn continuous non-credibility scores from stylistic features and dictionary features~\cite{przybyla2020capturing,baly2018predicting,baly-etal-2019-multi}.

\textbf{LLM extraction of credibility signals for weakly supervised authenticity prediction}: Several studies have employed large language models to convert expert-defined credibility indicators into usable signals under weak supervision~\cite{leite2025pastel,smith2022lml,yu-bach-2023-alfred,huang2024alchemist,hasanain-etal-2024-large,hamilton2024gptassisted}.
Leite et al.~\cite{leite2025pastel} proposed \textsc{PASTEL}, which uses zero-shot LLM extraction of credibility signals from full online articles. By contrast, \textsc{Tag2Cred} targets short, URL-sparse Telegram messages and uses a supervised, closed-vocabulary message-level codebook (theme, claim type, CTA, evidence) rather than open-ended article-level signals or source-identity features.
Modzelewski et al.~\cite{modzelewski-etal-2025-pcot} proposed PCoT, a method for zero-shot disinformation detection on online news and social media posts. PCoT first predicts high-level persuasion strategies with explanations and then conditions the final fake/real decision on that persuasion analysis; we therefore include it as a directly applicable content-only baseline for URL-masked Telegram messages.
Smith et al.~\cite{smith2022lml} used cue words as labeling functions in the weak supervision process to denoise and fuse multiple signals before training downstream classifiers. Srba et al. ~\cite{srba2024} provide a comprehensive survey of LLM-based credibility signal extraction methods that target long-form web content, which is a setting distinct from the short, URL-sparse Telegram messages addressed here. \citet{nikolaidis2024} confirms that persuasion-technique features are strongly discriminative for misinformation classification, consistent with our CTA and claim-framing ablations.
\section{Method}
\label{sec:method}

\subsection{\textsc{Tag2Cred} Framework Overview}
\label{sec:framework}
\begin{figure*}
    \centering
    \includegraphics[width=\textwidth]{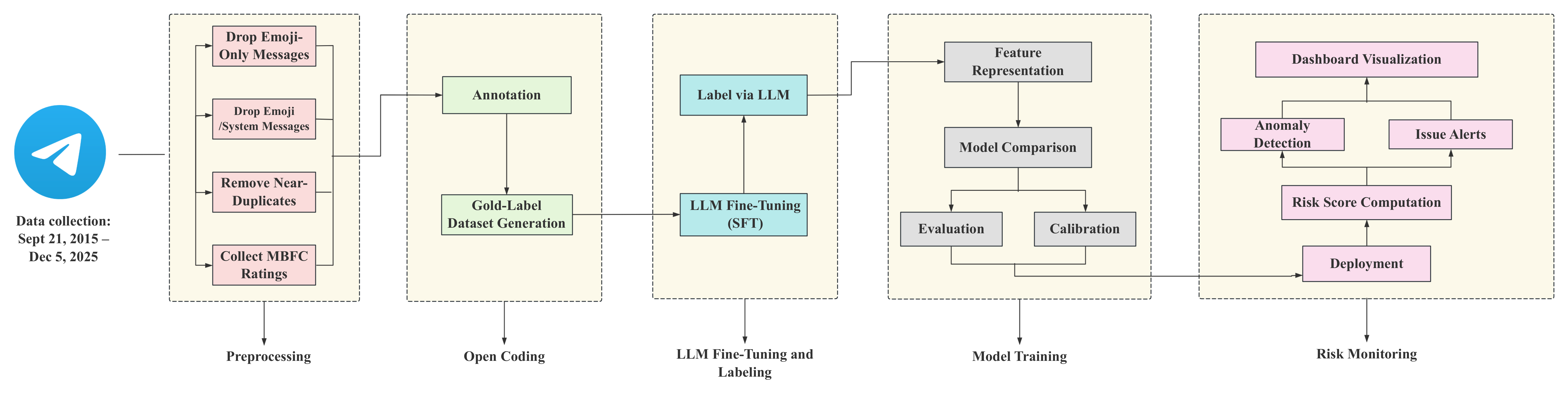}
    \caption{\textsc{Tag2Cred} pipeline: data collection and preprocessing; labeling; model training; risk scoring; and monitoring.}
    \label{fig:pipeline}
\end{figure*}

\textsc{Tag2Cred's} end-to-end workflow is summarized in Figure~\ref{fig:pipeline}. The workflow has five stages: (1) collect and preprocess messages and MBFC ratings; (2) use open coding to create a human-annotated gold-label dataset; (3) fine-tune Qwen-32B with supervised fine-tuning (SFT), verify it with a separate quality-control check, and use it to label data at scale; (4) construct the feature representation and run model comparison, including evaluation and calibration; and (5) deploy the selected model to compute risk scores and run the monitoring case study.

\subsection{Data Collection and Preprocessing}
\label{sec:data-preprocess}
Our data collection spanned from September 21st, 2015, to December 5th, 2025. We began with public seed channels and then expanded through platform-native forwarding and cross-promotion links. We collected messages from public Telegram channels with a minimum subscriber count of 500 using the Telethon Python client together with Telegram's MTProto API~\cite{telethon}. We were able to collect 9,371,099 messages from 234 channels.
We then apply an auditable cleaning and near-deduplication pipeline and extract, resolve, and canonicalize URLs to registered domains.
The supervised experiments use an MBFC-linkable subset~\cite{mbfc} of 87,936 cleaned messages: messages that passed preprocessing and contain at least one URL whose canonical domain appears in our MBFC dump. After thresholding, 65,933 high-confidence labeled messages remain for supervised training and evaluation. We retain the canonical domain field for supervision construction and domain-disjoint evaluation splits.

\begin{figure}[t]
    \centering
    \includegraphics[width=\linewidth]{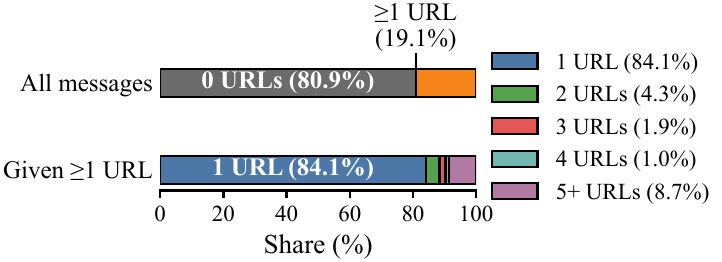}
    \caption{URL sparsity in the full corpus. Most messages contain no explicit web URL. Among messages with at least one URL, 84.1\% contain exactly one.
}
    \label{fig:url-stats}
\end{figure}

\textbf{URL sparsity and MBFC coverage.}
Telegram has relatively few explicit web URLs, as shown in Figure~\ref{fig:url-stats}. The large corpus contains 1,788,007 messages (19.1\%) with at least one explicit web URL and 7,583,092 messages (80.9\%) with none. Matching extracted URL domains with MBFC using the raw data yields 431,900 messages with at least one MBFC-rated domain (24.2\% of URL-bearing messages; 4.61\% overall).

\begin{figure}[t]
  \centering
  \subfloat[Credibility\label{fig:mbfc-cred}]{%
  \hspace*{-2em}
    \includegraphics[width=0.42\columnwidth]{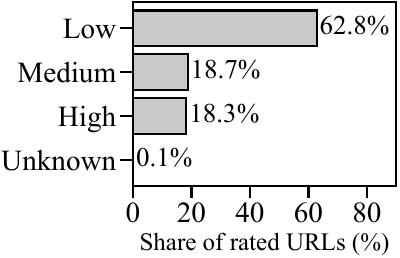}%
  }\hfill
  \subfloat[Factuality\label{fig:mbfc-fact}]{%
    \makebox[0.5\columnwidth][c]{%
      \hspace*{-3em}
      \includegraphics[width=0.48\columnwidth]{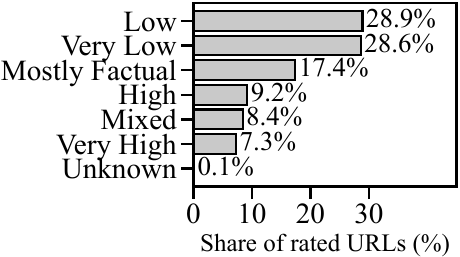}%
    }%
  }
  \caption{MBFC Credibility and Factuality Distributions for Rated URLs.}
  \label{fig:mbfc-profile}
\end{figure}

\textbf{MBFC profile of rated URLs.}
The MBFC profile of rated URLs shows that credibility is largely Low (62.8\%), with Medium (18.7\%) and High (18.3\%). Factuality falls mostly into Very Low (28.6\%) and Low (28.9\%) (Figure~\ref{fig:mbfc-profile}).

\textbf{Auditable cleaning and near-duplicate removal.}
Following deterministic normalization and removal of non-labelable or low-information posts, we perform single-pass greedy cross-channel clustering. We first merge exact fingerprints based on canonical text, token-set digest, and 3-gram shingle digest. We then merge candidates when any of the following conditions hold: MinHash-Jaccard $\ge 0.85$ on 3-token shingles (LSH), SimHash Hamming $\le 3$ on 5-gram characters, or embedding cosine $\ge 0.95$ with all-MiniLM-L12-v2~\cite{reimers-gurevych-2019-sentence}.
We remove forwarded-message exact duplicates by (fwd\_id, text).

\textbf{URL resolution, domain canonicalization, and URL extraction.} For each cleaned message, we extract the explicit web URLs, resolve common redirection patterns, and map the final URL to a canonical registered domain. Processing methods include lowercasing hostnames, removing the `www.' prefix, and simplifying based on public-suffix rules~\footnote{\url{https://publicsuffix.org/}}. Finally, we use the resulting domain identifier to match MBFC and serve as a grouping variable for domain‑disjoint evaluation.

\textbf{URL masking.}
We apply URL masking across all models before feature extraction to prevent source-identity leakage. Canonical domains are retained only as metadata for supervision and splitting, never as model features.

\subsection{Groundtruth Creation and Feature Construction}
\label{sec:groundtruth}

We construct three complementary forms of ground truth: (i) MBFC-derived distant supervision for credibility risk, (ii) a human-coded rhetorical tag codebook for training LLM taggers, and (iii) manually annotated binary credibility-risk labels for the native zero-URL benchmark.

\subsubsection{MBFC-Derived Distant Supervision for Risk.}
\label{sec:mbfc-supervision}

MBFC provides domain-level categorical ratings, which we convert into continuous risk scores $R_{\mathrm{msg}} \in [0,1]$ and high-confidence binary labels $y \in \{0,1\}$ for supervised experiments. This module consists of three steps:
\textbf{(1) Raw MBFC ratings}.
Using Rapid API~\footnote{\url{https://rapidapi.com/mbfcnews/api/media-bias-fact-check-ratings-api2}}, for each domain that matches MBFC, we obtain two categorical ratings, a credibility rating and a factual reporting rating, while retaining auxiliary metadata such as bias, media type, and country for descriptive analysis. Risk mapping only uses the first two ratings. \textbf{(2) Domain-level canonicalization.} Due to aliases, redirects, and multiple matches, MBFC joins may result in duplicate or conflicting label pairs for the same domain. We first standardize the label strings and then merge the candidate matches into a unique label pair for each normalized domain by selecting the most frequent (credibility, factual-reporting) combination. If a tie occurs, we prioritize selecting labels with greater credibility and greater factual reporting. \textbf{(3) Category-to-risk mapping.} We convert MBFC categories into two normalized risk components on the unit interval via ordinal scaling. For credibility, we transform labels High/Medium/Low to 0/1/2 and compute $r_C(d)=\mathrm{rank}_C(d)/2$, yielding $r_C(d)\in\{0,0.5,1\}$. For factual reporting, we convert Very High/High/Mostly Factual/Mixed/Low/Very Low to 0/1/2/3/4/5 and compute $r_F(d)=\mathrm{rank}_F(d)/5$, yielding $r_F(d)\in\{0,0.2,0.4,0.6,0.8,1\}$.
We then define the domain-level risk score:
\[
R(d) \;=\; 1 - (1-r_C(d))(1-r_F(d)),
\]
where a larger $R(d)$ indicates lower expected credibility.
Let $\mathcal{D}(m)$ be the set of resolved URLs in message $m$ that match MBFC entries. We define message-level risk as:
\[
R_{\mathrm{msg}}(m)=\max_{d\in \mathcal{D}(m)} R(d)
\]
This conservative rule uses the worst-linked source. Messages with $\mathcal{D}(m)=\varnothing$ remain unlabeled under this supervision scheme. For domain-disjoint evaluation, we assign each message a supervising domain:
\[
d(m)=\arg\max_{d\in\mathcal{D}(m)} R(d)
\]
Ties are broken deterministically, and group splits are performed on $d(m)$.

We label $y(m)=0$ if $R_{\mathrm{msg}}(m)\le \tau_{\mathrm{low}}$, label $y(m)=1$ if $R_{\mathrm{msg}}(m)\ge \tau_{\mathrm{high}}$, and leave messages unlabeled otherwise. We evaluated four threshold pairs. Among the four, $(\tau_{\mathrm{low}},\tau_{\mathrm{high}}) = (0.3, 0.8)$ achieves the best Macro-F1$\times$domain-coverage trade-off (TF/\textsc{Tag2Cred}/TF+\textsc{Tag2Cred}: 0.61/0.62/0.63) and is used throughout. Messages with $R_{\mathrm{msg}}(m)\in(\tau_{\mathrm{low}},\tau_{\mathrm{high}})$ are treated as fuzzy samples and excluded from supervised training and evaluation.
Since MBFC-derived distant supervision relies on URLs, it cannot label native zero-URL messages. For the strictly deduplicated native zero-URL benchmark of 1,252 messages, two independent coders manually assigned binary high-risk/low-risk credibility labels; inter-coder agreement was $\kappa=0.81$, and disagreements were resolved by consensus.

\begin{table*}[t]
\begingroup
\centering
\small
\setlength{\tabcolsep}{4pt}
\renewcommand{\arraystretch}{1.15}
\caption{Agreement (F1) between candidate LLM taggers and the human-coded rhetorical tag ground truth across the four codebook fields (based on 273 human-agreed gold samples).}

\begin{tabularx}{\textwidth}{@{} r L r r r r r @{}}
\toprule
\multicolumn{1}{c}{\#} &
\multicolumn{1}{c}{Model} &
\multicolumn{1}{c}{Theme} &
\multicolumn{1}{c}{\shortstack{Claim type /\\Framing}} &
\multicolumn{1}{c}{\shortstack{Call-to-\\Action}} &
\multicolumn{1}{c}{\shortstack{Evidence/Support\\Shown}} &
\multicolumn{1}{c}{Overall} \\
\midrule

1  & ChatGPT 5 (effort/verbosity) Medium/Medium & 87.31\% & 73.95\% & \textbf{89.13\%} & 80.59\% & 82.75\% \\
2  & Claude Sonnet 4                            & \textbf{89.02\%} & 68.73\% & 87.48\% & 79.15\% & 81.10\% \\
3  & Claude Opus 4.1                            & 84.69\% & 70.65\% & 85.29\% & 76.46\% & 79.27\% \\
4  & DeepSeek                                   & 85.31\% & 63.98\% & 84.60\% & 79.86\% & 78.44\% \\
5  & ChatGPT 5 (effort/verbosity) High/High     & 74.53\% & 65.44\% & 72.46\% & 70.48\% & 70.73\% \\
6  & ChatGPT 5 (effort/verbosity) High/Medium   & 74.09\% & 62.65\% & 73.64\% & 66.40\% & 69.20\% \\
7  & Qwen-32b-SFT                               & 83.18\% & \textbf{83.27\%} & 83.46\% & \textbf{83.92\%} & \textbf{83.46\%} \\
8  & ChatGPT 4o                                 & 82.98\% & 54.16\% & 70.83\% & 57.92\% & 66.47\% \\
9  & GPT-OSS-120B                               & 69.46\% & 53.00\% & 73.89\% & 66.21\% & 65.64\% \\
10 & Llama 3.3 (sub: llama3.1:8b)               & 82.08\% & 39.41\% & 53.51\% & 46.71\% & 55.43\% \\
\bottomrule
\end{tabularx}
\label{tab:llm_tagger_f1}
\endgroup
\end{table*}

\subsubsection{Rhetorical Tag Codebook and Gold Labels.}
\label{sec:human-tags}
We define a closed-vocabulary rhetorical tag schema with four fields:
\textit{Theme} (11 labels), \textit{Claim type / framing} (11 labels),
\textit{Call to action (CTA)} (7 labels), and \textit{Evidence / support shown} (6 labels).
The complete coding manual specifies decision order (Q1$\rightarrow$Q4), a
``51\%'' inclusion rule, tie-breakers, and field constraints (see ``Codebook Prompt'' in the appendix).
Using an iterative open-coding procedure, two coders independently annotated 833 messages. For inconsistent results, coders discussed how to resolve disagreements. To assess the inter-coder reliability, we performed a Cohen-Kappa test. The Kappa score was 0.79, indicating substantial agreement.
Table~\ref{tab:example-messages} provides representative messages and their gold labels.

\subsubsection{LLM-Assisted Tag Assignment.}
\label{sec:llm-tagging}
We evaluated several candidate taggers and selected a supervised, open-source model that we fine-tuned on our codebook. The tagger was trained to produce four fixed fields in this order: Theme, Claim type/framing, CTA, and Evidence/Support.
We manually labeled 833 messages in total; 560 were used for supervised fine-tuning, and the remaining 273 were held out as an evaluation set.
Fine-tuning uses two epochs, LoRA r=32 $\alpha$=64,
batch size 2 with 8-step gradient accumulation, lr $2\times10^{-4}$, max length 4096, with assistant-only loss masking. At inference, we decode greedily (\texttt{temperature}=0). Theme (Q1) is treated as a single-label field, while Claim Type (Q2), CTA (Q3),
and Evidence (Q4) are multi-label fields whose
outputs may contain multiple labels. The outputs are normalized and converted into sparse bucketed feature vectors downstream. Table~\ref{tab:llm_tagger_f1} shows the results.
Qwen-32b-SFT achieves the best overall agreement (83.46\% mean field F1) and is consistently strong across all four fields. Several general-purpose chat models show uneven field-level performance (e.g., collapsing on CTA in some configurations, not outputting JSON at all), and open-weight Llama baselines often violate the closed-vocabulary constraints, yielding near-zero field F1.

\section{Classification and Evaluation}
\label{sec:features}

For the MBFC-linkable supervised subset, we assign a binary credibility-risk label $y(m) \in \{0,1\}$ using the distant-supervision rule above; the native zero-URL benchmark uses the manual binary labels described above.
The predicted probability $\hat{p}(y=1\mid m)\in[0,1]$ serves as the message-level credibility-risk score.

To verify the generalization ability of the model, we design two repeated group-disjoint evaluation families related to actual deployment scenarios to avoid information leakage caused by inter-group overlap.
(1) \textbf{Domain-disjoint partitioning}: the grouping key is the supervising domain $d^{(m)}$ associated with the MBFC-derived distant-supervision rule. For each seed, we generate 2048 candidate splits and select the one minimizing a balance objective over held-out row share, prevalence deviation, class-specific share, group ratio, and minority-class shortfall.
(2) \textbf{Channel-disjoint partitioning}: the same balanced grouped-split search is used with channel identifiers as the grouping key. For both families, approximately one-fifth of the data is held out in each outer split, and final results are reported as mean $\pm$ standard deviation over 50 frozen grouped outer splits.

\begin{table*}[t]
\centering
\caption{Performance on the strictly deduplicated native zero-URL set of 1,252 messages with manually annotated binary credibility-risk labels. Labels were assigned by two independent coders ($\kappa=0.81$). Values are means across 10 channel-disjoint folds. Subscripts give standard deviations.}
{\normalsize
\begin{tabular}{lccccc}
\toprule
& \multicolumn{5}{c}{10-fold channel-disjoint CV} \\
Model & Acc. & AUC & Macro-F1 & Brier & ECE \\
\midrule
TF--IDF & 0.639\,\textsubscript{0.076} & 0.644\,\textsubscript{0.095} & 0.593\,\textsubscript{0.087} & 0.220\,\textsubscript{0.026} & 0.139\,\textsubscript{0.052} \\
SBERT & 0.668\,\textsubscript{0.081} & 0.714\,\textsubscript{0.105} & 0.635\,\textsubscript{0.088} & 0.238\,\textsubscript{0.073} & 0.214\,\textsubscript{0.072} \\
\textsc{Tag2Cred} & 0.919\,\textsubscript{0.038} & \textbf{0.979}\,\textsubscript{0.007} & 0.912\,\textsubscript{0.036} & \textbf{0.060}\,\textsubscript{0.017} & \textbf{0.077}\,\textsubscript{0.027} \\
Ensemble (TF--IDF + \textsc{Tag2Cred}) & \textbf{0.923}\,\textsubscript{0.035} & 0.978\,\textsubscript{0.007} & \textbf{0.914}\,\textsubscript{0.036} & 0.063\,\textsubscript{0.017} & 0.111\,\textsubscript{0.024} \\
Ensemble (TF--IDF + \textsc{Tag2Cred} + SBERT) & 0.922\,\textsubscript{0.029} & 0.977\,\textsubscript{0.011} & 0.913\,\textsubscript{0.026} & 0.065\,\textsubscript{0.021} & 0.101\,\textsubscript{0.023} \\
LLM-zero-shot (GPT-5.4-mini) & 0.836\,\textsubscript{0.045} & 0.910\,\textsubscript{0.032} & 0.822\,\textsubscript{0.045} & 0.173\,\textsubscript{0.075} & 0.180\,\textsubscript{0.101} \\
PASTEL & 0.744\,\textsubscript{0.068} & 0.836\,\textsubscript{0.053} & 0.715\,\textsubscript{0.076} & 0.228\,\textsubscript{0.060} & 0.235\,\textsubscript{0.065} \\
PCoT & 0.737\,\textsubscript{0.071} & 0.753\,\textsubscript{0.081} & 0.685\,\textsubscript{0.071} & 0.200\,\textsubscript{0.030} & 0.148\,\textsubscript{0.061} \\
ARG-D-style & 0.658\,\textsubscript{0.140} & 0.733\,\textsubscript{0.086} & 0.585\,\textsubscript{0.136} & 0.215\,\textsubscript{0.064} & 0.190\,\textsubscript{0.089} \\
\bottomrule
\end{tabular}
}
\label{tab:zero_url_native}
\end{table*}

\textbf{Supervised models (feature representations).}
We trained $\ell_2$-regularized logistic regression classifiers on three URL-independent feature sets: term frequency--inverse document frequency (TF--IDF) features~\cite{salton1988tfidf}, \textsc{Tag2Cred} tag vectors, and Sentence-BERT (SBERT) embeddings~\cite{reimers-gurevych-2019-sentence}.
In the 65,933-message grouped evaluation, TF--IDF and SBERT used balanced class weights. For \textsc{Tag2Cred}, the inner group-aware cross-validation compared balanced and unweighted fits.
The grouped evaluation used scikit-learn logistic regression on the SBERT embeddings~\cite{pedregosa2011sklearn}.

\textbf{Weighted ensembles.}
For the 65,933-message grouped evaluation, we averaged the base-model probabilities for two ensembles, TF--IDF+\textsc{Tag2Cred} and TF--IDF+\textsc{Tag2Cred}+SBERT. For each outer split, we selected blend weights and a decision threshold using macro-F1 on group-disjoint out-of-fold predictions from the training partition.

\textbf{Training configuration.}
Hyperparameters and decision thresholds are selected via 5-fold group-aware cross-validation on the outer training partition. The final model is refit on the full training partition and evaluated once on the held-out test partition.

\begin{table*}[t]
  \centering
  \setlength{\tabcolsep}{3.0pt}
  \renewcommand{\arraystretch}{1.06}
  \footnotesize
\caption{Representative Telegram messages with gold labels for the four-field rhetorical codebook (Q1--Q4).}
  \begin{tabularx}{\textwidth}{@{}c L >{\raggedright\arraybackslash}p{0.12\textwidth} >{\raggedright\arraybackslash}p{0.19\textwidth} >{\raggedright\arraybackslash}p{0.13\textwidth} >{\raggedright\arraybackslash}p{0.12\textwidth}@{}}
    \toprule
\textbf{\#} & \textbf{Message} & \textbf{Theme (Q1)} & \textbf{Claim type (Q2)} & \textbf{CTA (Q3)} & \textbf{Evidence (Q4)} \\
    \midrule

    1 &
    \$25,000 VALUE GIVEAWAY + 2000 IGGYBOY WHITELIST SPOTS PRIZE POOL 150 IggyBoy NFT Price Cashback 5000 USDT Prize Pool 2000 IggyBoy Whitelist Spots HOW TO PARTICIPATE Complete all mandatory gleam tasks: [URL] Ends Feb 21, 12PM CET
    &
    Finance/\allowbreak Crypto
    &
    Scarcity/\allowbreak FOMO tactic\newline Verifiable factual statement
    &
    Join/\allowbreak Subscribe\newline Visit external link /\ watch video
    &
    Link/\allowbreak URL\newline Statistics
    \\

    2 &
    \$ETHUSDT Just sending you free money here. Trade closed. [URL]
    &
    Finance/\allowbreak Crypto
    &
    Promotional hype /\ exaggerated profit guarantee
    &
    Buy /\ invest /\ donate
    &
    Link/\allowbreak URL\newline Chart /\ price graph /\ TA diagram
    \\

    3 &
    A buried EPA report linking glyphosate to non-Hodgkin lymphoma has recently been exposed. [URL]
    &
    Public health \& medicine
    &
    Rumour /\ unverified report
    &
    Visit external link /\ watch video
    &
    Link/\allowbreak URL
    \\

    \bottomrule
  \end{tabularx}
  \label{tab:example-messages}
\end{table*}

\textbf{Probability calibration.}
For the grouped evaluation, we assess calibration on each held-out test partition using the probabilities from the fitted classifiers or their weighted averages.
We compute the Brier score as $n^{-1}\sum_{i=1}^{n}(p_i-y_i)^2$, where $p_i$ is the predicted high-risk probability and $y_i$ is the binary label for test message $i$~\cite{brier1950brierscore}. For expected calibration error (ECE), we partition $[0,1]$ into equal-width bins and weight the absolute difference between each bin's mean predicted probability and observed positive rate by its share of test messages~\cite{guo2017calibration}. The grouped TF--IDF, SBERT, \textsc{Tag2Cred}, and ensemble evaluations use 10 bins. The ARG-D-style evaluation uses 15.

\section{Results: Representation comparisons under URL masking}
We evaluated three URL-independent supervised representation methods and three LLM-based baseline models.
Table~\ref{tab:perf} shows the results.

\textbf{The vocabulary baseline model (URL-masked TF--IDF).}
TF-IDF captures surface vocabulary patterns while URL masking prevents source-identity leakage. Under domain-disjoint evaluation, it achieved an AUC of 0.848 $\pm$ 0.016, Macro-F1 0.761 $\pm$ 0.016, and ECE 0.110 $\pm$ 0.011.

\textbf{The semantic embedding baseline model (URL-masked SBERT + logistic regression).}
Under URL masking with a frozen encoder and logistic regression, SBERT achieves AUC 0.833 $\pm$ 0.008, Macro-F1 0.756 $\pm$ 0.007, ECE 0.156 $\pm$ 0.011.
In this repeated grouped-split re-evaluation, the ranking is unchanged: \textsc{Tag2Cred} outperforms TF--IDF and SBERT as a single-view URL-independent model. Table~\ref{tab:sig_combined} in the appendix reports the Holm-corrected Wilcoxon comparisons for both split families.

\textbf{\textsc{Tag2Cred} feature (rhetorical tag representation).}
\textsc{Tag2Cred} maps four rhetorical tag fields (Theme, Claim type/framing, Call-to-Action, Evidence/Support) into a sparse closed-vocabulary feature space. {{Our main representation contains per-field bucket indicators together with structural tokens for field cardinality, missingness, and within-field multi-label combinations, and it adds selected grouped cross-field pairings (including Claim-CTA, Claim-Evidence, and Evidence-CTA). For example, we could introduce new pairs such as \textit{Speculative forecast/prediction}, \textit{Share/repost/like},
\textit{Promotional hype/exaggerated profit guarantee}, \textit{Visit external link/watch video}, \textit{Opinion/subjective statement}, \textit{Chart/price graph/TA diagram}, and \textit{Statistics}, \textit{Share/repost/like}. These would be added directly after the single one-hot feature vector for training. }} Although the model ignores word/lexical-level text, it still achieves competitive performance, with a macro-F1 value of {{$0.821\pm0.011$.}} The Brier score is {{$0.153\pm0.007$}}, and the ECE is {{$0.124\pm0.019$}}, which is pivotal for risk-weighted monitoring tasks. Because the resulting feature space remains sparse and interpretable, the learned risk scores can be analyzed directly through linear coefficients. The ablations below show that \textit{Call-to-Action} (CTA) is the strongest single cue, but the main lift comes from claim framing and selected grouped cross-field structure rather than from topical theme alone. This interpretability makes the representation a core method in the monitoring scenarios described below.

\textbf{Composite representation (weighted ensembles).}
The weighted ensemble further improves performance, and the complete ensemble scheme integrating TF--IDF, \textsc{Tag2Cred}, and SBERT achieves strong overall performance, with AUC reaching $0.893\pm0.019$, macro-F1 value of $0.828\pm0.013$, Brier score $0.143\pm0.007$, and ECE of $0.113\pm0.013$ (Table~\ref{tab:perf}).

\begin{table}[t]
\centering
\caption{Grouped evaluation on 65,933 messages. Values are means across 50 repeated grouped splits, with standard deviations in subscript. Bold marks the best mean per metric within each split type. TF = TF--IDF and T2C = \textsc{Tag2Cred}. The zero-shot model is GPT-5.4-mini.}

{\fontsize{9}{10}\selectfont
\setlength{\tabcolsep}{2.5pt}
\renewcommand{\arraystretch}{1.06}

\begin{tabular*}{\columnwidth}{@{\extracolsep{\fill}}lrrrr@{}}
\hline
Model & Macro-F1 & AUC & Brier & ECE \\
\hline

\multicolumn{5}{@{}l}{\textit{Domain-disjoint}} \\
\hline
TF & .761\,\textsubscript{.016} & .848\,\textsubscript{.016} & .174\,\textsubscript{.006} & .110\,\textsubscript{.011} \\
SBERT & .756\,\textsubscript{.007} & .833\,\textsubscript{.008} & .197\,\textsubscript{.006} & .156\,\textsubscript{.011} \\
T2C & .821\,\textsubscript{.011} & .864\,\textsubscript{.014} & .153\,\textsubscript{.007} & .124\,\textsubscript{.019} \\
TF+T2C & .825\,\textsubscript{.013} & .880\,\textsubscript{.020} & .147\,\textsubscript{.008} & .115\,\textsubscript{.015} \\
TF+T2C+SBERT
& \textbf{.828}\,\textsubscript{.013} & \textbf{.893}\,\textsubscript{.019}
& \textbf{.143}\,\textsubscript{.007} & .113\,\textsubscript{.013} \\
Zero-shot
& .650\,\textsubscript{.016} & .739\,\textsubscript{.025} & .226\,\textsubscript{.013} & .150\,\textsubscript{.014} \\
PASTEL & .653\,\textsubscript{.014} & .746\,\textsubscript{.017} & .326\,\textsubscript{.014} & .321\,\textsubscript{.014} \\
PCoT & .467\,\textsubscript{.010} & .668\,\textsubscript{.034} & .227\,\textsubscript{.012} & \textbf{.087}\,\textsubscript{.017} \\
ARG-D-style & .758\,\textsubscript{.089} & .885\,\textsubscript{.033} & .276\,\textsubscript{.093} & .285\,\textsubscript{.118} \\
\hline
\multicolumn{5}{@{}l}{\textit{Channel-disjoint}} \\
\hline
TF & .659\,\textsubscript{.025} & .772\,\textsubscript{.025} & .185\,\textsubscript{.010} & .086\,\textsubscript{.018} \\
SBERT & .722\,\textsubscript{.014} & .805\,\textsubscript{.016} & .176\,\textsubscript{.009} & .101\,\textsubscript{.012} \\
T2C & .773\,\textsubscript{.013} & .811\,\textsubscript{.017} & .168\,\textsubscript{.010} & .104\,\textsubscript{.013} \\
TF+T2C & .774\,\textsubscript{.012} & .816\,\textsubscript{.017} & .162\,\textsubscript{.010} & .085\,\textsubscript{.025} \\
TF+T2C+SBERT
& \textbf{.782}\,\textsubscript{.013} & \textbf{.830}\,\textsubscript{.018}
& \textbf{.152}\,\textsubscript{.011} & \textbf{.077}\,\textsubscript{.017} \\
Zero-shot
& .662\,\textsubscript{.014} & .726\,\textsubscript{.020} & .263\,\textsubscript{.012} & .239\,\textsubscript{.026} \\
PASTEL & .595\,\textsubscript{.013} & .650\,\textsubscript{.016} & .356\,\textsubscript{.019} & .350\,\textsubscript{.018} \\
PCoT & .632\,\textsubscript{.017} & .665\,\textsubscript{.025} & .257\,\textsubscript{.012} & .194\,\textsubscript{.030} \\
ARG-D-style & .671\,\textsubscript{.125} & .798\,\textsubscript{.068} & .272\,\textsubscript{.077} & .271\,\textsubscript{.096} \\
\hline
\end{tabular*}
}

\label{tab:perf}
\end{table}

\textbf{Zero-shot baseline of large language model (direct risk score).}
Without supervised training, the large language model (GPT-5.4-mini) outputs a credibility-risk probability value $\hat{p}(y=1\mid m)\in[0,1]$ based on the MBFC standard with a zero-shot baseline (direct risk score).
To make a fair comparison with the supervised models, we tune only the decision threshold on train-side validation predictions. On the grouped URL-masked benchmark, the zero-shot baseline reaches AUC $0.739\pm0.025$ and Macro-F1 $0.650\pm0.016$ under domain-disjoint evaluation, and AUC $0.726\pm0.020$ and Macro-F1 $0.662\pm0.014$ under channel-disjoint evaluation; on the native zero-URL benchmark, it reaches AUC $0.910\pm0.032$ and Macro-F1 $0.822\pm0.045$ (Tables~\ref{tab:zero_url_native} \&~\ref{tab:perf}).

\textbf{PCoT baseline.} We evaluate PCoT~\cite{modzelewski-etal-2025-pcot}, a two-stage zero-shot disinformation detector. First, the model predicts high-level persuasion strategies and explains them. Then it uses that analysis to decide whether text is fake or real. We keep this official two-stage setup on the same URL-masked Telegram text and map fake/real outputs to high-risk/low-risk. PCoT performed worst in the domain-disjoint evaluation, with Macro-F1 0.467 and AUC 0.668, suggesting that its persuasion-conditioned pipeline struggles with short, URL-sparse Telegram messages (Tables~\ref{tab:zero_url_native} and~\ref{tab:perf}).

\textbf{ARG-D-style baseline.} We adapt the rationale-guided ARG/ARG-D framework~\cite{hu-etal-2024-bad-actor} to URL-masked Telegram messages. We train a rationale-aware teacher using textual-description and commonsense rationales, then distill a rationale-free student. We use frozen, mean-pooled \texttt{bert-base-uncased} embeddings. This allows us to scale grouped evaluation.  Domain-disjoint has AUC 0.885, Macro-F1 0.758 and Channel-disjoint got AUC 0.798, Macro-F1 0.671. Zero-URL: Accuracy 0.658, AUC 0.733, Macro-F1 0.585 (Tables~\ref{tab:zero_url_native} and~\ref{tab:perf}).

\textbf{The PASTEL-style baseline (LLM credibility signals + Snorkel LabelModel)} adapts the weak-signal pipeline of PASTEL style to the Telegram short message scenario~\cite{leite2025pastel}.
Specifically, the prompted LLM answers $K{=}19$ fixed article-style credibility-signal questions on URL-masked message text (e.g., lack of supporting evidence; presence of a call to action) and encodes \{Yes, No, Unsure\} as \{1, 0, -1\}, yielding a 19-dimensional weak-vote vector.
Under each external dataset partitioning scheme, only the training set voting matrix is used to fit the Snorkel LabelModel~\cite{ratner2017snorkel}, and the posterior probability $\hat{p}(y=1\mid m)$ output is threshold-tuned on the validation set.
Under this experimental setup, the PASTEL-style baseline is clearly weaker than \textsc{Tag2Cred} on both evaluation families. Under the domain-disjoint setting it achieves AUC $0.746\pm0.017$ and macro-F1 $0.653\pm0.014$ (Brier $0.326\pm0.014$, ECE $0.321\pm0.014$), while under the channel-disjoint setting it reaches AUC $0.650\pm0.016$ and macro-F1 $0.595\pm0.013$ (Brier $0.356\pm0.019$, ECE $0.350\pm0.018$) (Table~\ref{tab:perf}).

CTA-centered and grouped-feature ablations.

Table~\ref{tab:cta_grouped_ablations} shows that CTA is the strongest single field but does not explain everything. CTA-only yields Macro-F1 0.795 ± 0.033 and ROC-AUC 0.806 ± 0.029. Adding Theme reduces these to 0.785 ± 0.051 and 0.793 ± 0.026, whereas adding Evidence keeps Macro-F1 near 0.792 ± 0.015 and raises ROC-AUC to 0.822 ± 0.018. The greatest improvement comes from adding claim framing. Claim + CTA achieves Macro-F1 0.853 ± 0.024 and ROC-AUC 0.864 ± 0.019. Relative to CTA-only, this improves Macro-F1 by 0.058 and ROC-AUC by 0.058, and the Macro-F1 gain is significant (Holm-corrected Wilcoxon $p < 6.3 \times 10^{-15}$). In the full tag space, the grouped model performs slightly better than the flat all-tag model, achieving Macro-F1 0.821 ± 0.011 versus 0.815 ± 0.012 and ROC-AUC 0.864 ± 0.014 versus 0.862 ± 0.015. This shows cross-field pairings aids the model to learn more complicated signals. Theme and Evidence are overall weaker predictors but we retain them because they enable interpretable monitoring across topics and evidence types, at a cost of a 0.032 drop in Macro-F1.

\begin{table}[ht]
\centering
\footnotesize
\setlength{\tabcolsep}{3pt}
\renewcommand{\arraystretch}{1.05}
\caption{CTA-centered and grouped-feature ablations on 65{,}933 messages (mean$\pm$std over 50 repeated domain-disjoint grouped splits). Flat models use per-field bucket, count, missingness, and within-field combination tokens; grouped models add selected cross-field interaction tokens.}
\label{tab:cta_grouped_ablations}
\begin{tabularx}{\columnwidth}{@{}>{\raggedright\arraybackslash}Xcc@{}}
\toprule
\textbf{Model} & \textbf{Macro-F1} & \textbf{AUC} \\
\midrule
CTA only & $0.795\pm0.033$ & $0.806\pm0.029$ \\
No CTA (flat) & $0.737\pm0.016$ & $0.786\pm0.018$ \\
CTA + Theme & $0.785\pm0.051$ & $0.793\pm0.026$ \\
CTA + Evidence & $0.792\pm0.015$ & $0.822\pm0.018$ \\
CTA + Claim & $0.853\pm0.024$ & $0.864\pm0.019$ \\

Claim + CTA + Evidence (grouped) & $0.823\pm0.011$ & $0.868\pm0.012$ \\
All tags (flat) & $0.815\pm0.012$ & $0.862\pm0.015$ \\
All tags (grouped) & $0.821\pm0.011$ & $0.864\pm0.014$ \\
\bottomrule
\end{tabularx}
\end{table}
{
\textbf{Robustness slices for non-finance same-main splits.}
Telegram is full of crypto scams, so we test whether \textsc{Tag2Cred} still performs well after removing Finance/Crypto messages. Tables~\ref{tab:nonfinance_slice_nonfinance} and~\ref{tab:nonfinance_slice_nonchat} in the Appendix provide the full results. For brevity, we summarize the key findings here.
On the non-finance slice, \textsc{Tag2Cred} reaches Macro-F1 $0.815\pm0.012$, ROC--AUC $0.857\pm0.015$, PR--AUC $0.852\pm0.019$, and Brier $0.159\pm0.007$, outperforming TF--IDF ($0.757\pm0.018$ / $0.845\pm0.019$ / $0.830\pm0.026$ / $0.182\pm0.007$) and SBERT ($0.751\pm0.007$ / $0.830\pm0.009$ / $0.818\pm0.012$ / $0.208\pm0.006$). Under the stricter non-finance/non-chat slice, \textsc{Tag2Cred} remains strong (Macro-F1 $0.802\pm0.013$, ROC--AUC $0.840\pm0.020$, PR--AUC $0.820\pm0.030$, Brier $0.169\pm0.009$) and continues to outperform SBERT, while TF--IDF retains a small ROC--AUC edge over the tag-only model. The full ensemble remains strongest overall on both slices, indicating that rhetorical, lexical, and semantic views remain complementary outside crypto rather than reducing to a single crypto-CTA heuristic. Paired significance tests are provided in Tables~\ref{tab:nonfinance_sig_a} and~\ref{tab:nonfinance_sig_b}, and curated non-crypto high-risk examples are provided in Table~\ref{tab:nonfinance_examples} in the appendix.
}

\textbf{Tagger-noise stress test.}
To measure how sensitive our system is, we inject independent bit-flip noise into the \textsc{Tag2Cred} tag vectors and evaluate the resulting degradation.
Macro-F1 / AUC / Acc. degrade smoothly from $0.821\pm0.011$ / $0.864\pm0.014$ / $0.823\pm0.012$ at 0\% noise to $0.786\pm0.013$ / $0.851\pm0.016$ / $0.786\pm0.013$ at 5\%, $0.765\pm0.014$ / $0.832\pm0.016$ / $0.765\pm0.014$ at 10\%, and $0.727\pm0.016$ / $0.788\pm0.017$ / $0.727\pm0.015$ at 20\%, indicating that the downstream model is not brittle to moderate tagging noise.

\textbf{Summary.}
Combining \textsc{Tag2Cred} with lexical (TF--IDF) and semantic (SBERT) views yields the strongest overall trade-off between discrimination and calibration (Table~\ref{tab:perf}), supporting its use as a calibrated message-level risk score for monitoring.
\section{Monitoring Use Case: Risk-Weighted Strategy Dynamics}
This section presents key findings about high-risk rhetorical strategies on Telegram using risk-weighted tag dynamics. A small number of recurring strategies dominate the high-risk tail, with finance/crypto promotion and scarcity narratives being the most enriched. Sharp temporal bursts can also be reliably detected through tag distribution monitoring.
The risk score $\hat p(m)$ is the model output based on \mbfc-based weak supervision.

\subsection{Monitoring setup}

\textbf{Monitoring window.}
We tag 229,276 messages from 115 channels (Jan 2024-Nov 2025), scoring each with the ensemble model while using tag representations for aggregation.

\textbf{Risk mass and risk share.}
We set $\mathcal{M}$ as a message set (for example, messages of a week, messages of a channel community, or messages corresponding to a subset of tags).
We define the risk mass as $\sum_{m\in\mathcal{M}} \hat p(m)$, which can be understood as the expected message count under the proxy label.
For the tag $t$ and its corresponding message set $\mathcal{M}_t$, we define:
\[
\text{RiskShare}(t) = \frac{\sum_{m\in\mathcal{M}_t} \hat p(m)}{\sum_{m\in\mathcal{M}} \hat p(m)},\quad
\text{VolShare}(t) = \frac{|\mathcal{M}_t|}{|\mathcal{M}|}.\]

\subsection{High-risk Tail and Enriched Strategies}

\textbf{High-risk tail.}
We define the top 5\% of messages by risk score $\hat p(m)$ in the monitoring window as the high-risk tail $H$.
This screening obtains a total of 11,464 messages, and the corresponding risk-score threshold is $\hat p(m) \ge 0.9818$.

\textbf{Enrichment analysis based on log-odds ratio with informative Dirichlet prior.}
In the case of sparse counts, lift ratios (enrichment multiples) may be unstable.
Therefore, we use the log-odds ratio with an informative Dirichlet prior~\cite{monroe2008} to quantify which tags in the high-risk tail $H$ are overrepresented compared with the non-high-risk message set $\mathcal{M}\setminus H$. We report $z$-scores (positive values indicate that the tag is enriched in the high-risk tail). Table~\ref{tab:logodds} reports the most enriched tags in the high-risk tail.
\begin{table}[t]
\centering
\small
\caption{Tags most enriched in the high-risk tail (top 5\%).}
\begin{tabular}{llr}
\toprule
\textbf{Field} & \textbf{Tag} & \textbf{$z$ (tail)} \\
\midrule
Theme & Finance/Crypto & 4.50 \\
Theme & Conversation/Chat/Other & 4.13 \\
Theme & Public health \& medicine & 2.24 \\
\midrule
Claim type & Scarcity/FOMO tactic & 5.22 \\
Claim type & Announcement & 4.83 \\
Claim type & Promotional hype / exaggerated profit & 4.62 \\
\midrule
CTA & Join/Subscribe & 3.60 \\
CTA & Buy/Invest/Donate & 3.03 \\
CTA & Engage / ask questions & 2.36 \\
\midrule
Evidence & Chart/price graph/TA diagram & 3.36 \\
Evidence & Statistics & 2.79 \\
Evidence & Quotes/Testimony & 1.68 \\
\bottomrule
\end{tabular}

\label{tab:logodds}
\end{table}

\begin{figure*}[t]
  \centering
  \subfloat[JS Divergence\label{fig:WeekA}]{%
    \includegraphics[width=.39\textwidth]{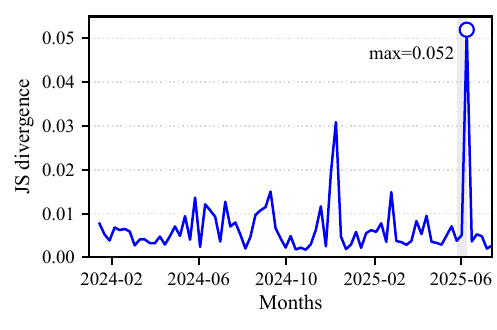}%
  }
  \subfloat[Weekly Share\label{fig:WeekB}]{%
    \includegraphics[width=.39\textwidth]{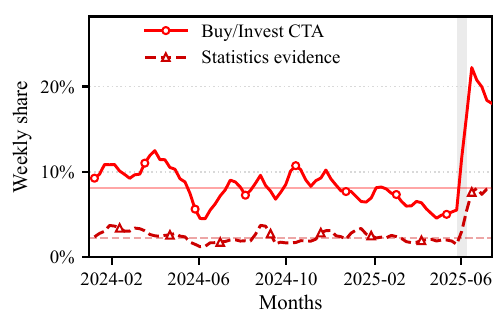}%
  }
  \caption{Week-over-week Jensen--Shannon divergence of the strategy-family distribution (a) and weekly message shares for the Buy/Invest CTA and Statistics evidence tags (b).}
  \label{fig:montly-shift}
\end{figure*}
\textit{Finance/Crypto} and \textit{Conversation/Chat/Other} account for 70.6\% of total messages, but contribute 74.0\% of the total risk mass. However, this preserved monitoring branch is not crypto-only. Excluding \textit{Finance/Crypto} still leaves 150,908 messages across 104 channels, and excluding both \textit{Finance/Crypto} and \textit{Conversation/Chat/Other} still leaves 67,285 messages across 93 channels (29.35\% of monitoring volume). This stricter non-finance/non-chat slice spans several substantive themes, including \textit{Technology} (38,810 messages across 55 channels), \textit{Gaming/Gambling} (6,316 across 35), \textit{Lifestyle \& well-being} (5,999 across 48), \textit{Politics} (5,305 across 67), and \textit{Public health \& medicine} (4,295 across 65), and it also persists in the high-risk tail, where it still contains 1,402 messages across 47 channels. In particular, \textit{Public health \& medicine} provides a clear non-crypto counterexample, with a high-tail lift of 1.48 despite its much smaller share of monitoring volume. Messages without a call to action dominate the total message volume (63\%), but are less prominent in the high-risk tail (lift is 0.69). The CTAs `Join/Subscribe' and `Buy/Invest/Donate' strongly correspond to the high‑risk tail, which is consistent with information mobilization and funneling behaviors.
Evidence-related features show an important pattern: Although `None/assertion only' makes up 84.7\% of the risk mass, messages with `Chart/price graph' evidence have a tail lift of 5.94 and nearly 30\% fall into the high‑risk tail.

\subsection{Prototypes and Strategy Families}

\textbf{Prototype definition.} A prototype is the unique combination of theme + claim type + CTA + evidence. The prototype provides an easy-to-understand ``strategy fingerprint'' for recurring message types.

\textbf{Coverage.} The top 10 prototypes cover 56.4\% of messages and 54.8\% of the high-risk tail; the top 50 cover 80.3\% and 76.5\% respectively, making monitoring tractable.

\textbf{Clustering into strategy families.}
To summarize prototypes at a higher level, we embed the prototypes based on TF--IDF over tag tokens, and then use the $k$-means algorithm for clustering.
We choose $k=25$ as the number of clusters to strike a balance between granularity and interpretability (Silhouette analysis \cite{rousseeuw1987} supports smaller values $k$, but produces overly coarse families).
The strategy families most related to the high-risk tail include: finance/crypto promotion with scarcity narratives and mobilizing CTAs (tail lift up to 4.91), and a small but distinctive strategy family related to public-health rumors with testimonial evidence.

\subsection{Temporal Dynamics: Drift and Bursts}

\textbf{Weekly drift (distribution shift).}
We calculated the week-over-week Jensen--Shannon (JS) divergence \cite{lin1991} of the 25 strategy-family distributions, used natural logarithms in the calculation process, and filtered out weeks with fewer than 200 tagged messages to ensure the stability of the results. Figure~\ref{fig:montly-shift} plots the week-over-week drift in the 25 strategy-family distribution.
The maximum weekly drift observed is 0.052 (between May 26--June 1, 2025, and June 2--June 8, 2025), indicating that the distribution of strategy families changed significantly.
This week saw a sharp increase in the ``Buy/Invest'' CTA (accounting for 23.5\%, while the median was 8.17\%) and a significant increase in the ``Statistics'' evidence (accounting for 6.73\%, while the median was 2.27\%).

\textbf{Peak-to-median ratio (sharp short-term peaks).}
For each tag, we define the peak-to-median ratio as $\max_w s_w / \mathrm{median}_w(s_w)$, where $s_w$ represents the weekly share of the tag.
Figure~\ref{fig:montly-shift}(b) shows that the Buy/Invest CTA peaked at 23.5\% during June 2--8, 2025 (median 8.17\%, peak-to-median ratio 2.88). Statistics evidence accounted for 6.73\% of messages that week and peaked at 9.39\% during July 7--13.

\textbf{Validation.}
Across all primary comparisons, paired Wilcoxon signed-rank tests over 50 repeated grouped splits confirm that \textsc{Tag2Cred}'s improvement over TF--IDF and SBERT are reliable under domain-disjoint and channel-disjoint evaluations (Table~\ref{tab:sig_combined}).
To assess sensitivity to tagging, we inject independent bit-flip noise into the \textsc{Tag2Cred} tag vector which confirms Macro-F1 degrades from $0.821$ at $0\%$ noise to $0.727$ at $20\%$.

\textbf{Robustness on non-English messages.} We evaluated relabeled \textsc{Tag2Cred} on 5,000 Russian and Farsi messages from public Telegram channels (Table~\ref{tab:ru-fa-stress}). Its macro-F1 exceeded the four baselines (zero-shot, PASTEL, PCoT, and ARG-D) on both domain-disjoint splits (0.580 versus 0.383--0.393) and channel-disjoint splits (0.571 versus 0.405--0.418). On domain-disjoint splits, \textsc{Tag2Cred} scored above TF--IDF (0.554) and below SBERT (0.627). On channel-disjoint splits, both TF--IDF (0.694) and SBERT (0.687) scored higher. These results point to the need for language-specific codebook validation and further tagger fine-tuning.

\textbf{Summary.}
Risk-weighted tag dynamics provide a compact monitoring layer in addition to message-level risk scores. In our monitoring window, the high-risk tail is disproportionately enriched for scarcity/promotional claim frames and mobilizing CTAs (Table~\ref{tab:logodds}). Prototype coverage remains concentrated. This should help analysts to focus on a small number of recurring strategies, while drift and burst metrics will surface weeks and tags that warrant deeper manual review (Figure~\ref{fig:montly-shift}). The Russian/Farsi stress test further shows that \textsc{Tag2Cred}'s structured rhetorical signal transfers meaningfully across languages and consistently outperforms all zero-shot baselines, with SBERT ahead on macro-F1 in both split families and TF--IDF ahead on channel-disjoint splits. All suggest that multilingual codebook validation and tagger fine-tuning remain important directions for future work.

\section{Limitations and Ethical Considerations}
The monitoring signal in this study is based on domain rating so that it may still inherit the limitations and deviations of external rating sources (MBFC).
Our 234 channels were selected with a 500-subscriber minimum, which may over-represent large, active channels and introduce topical skew toward finance/crypto content.
Although the tagger is highly consistent with the gold-label tags, it may still miss subtle contextual information, ironic expressions, or specific language cues.
The supervised study is limited to English-language channels. Non-English communities (e.g., Russian, Farsi) may exhibit rhetorical norms and misinformation patterns that the current codebook and tagger do not fully capture.

\textbf{Ethical Statement.}
We analyze only public channel content from the Telegram MT-Proto API and focus on aggregated patterns rather than attributing risks to individuals.

\textbf{Reproducibility.} Analysis code and aggregate results are available at \url{https://anonymous.4open.science/r/tag-signal-artifacts-BBAF/}

\section{Conclusion}

URL sparsity on Telegram renders link-based and lexical monitoring inadequate. \textsc{Tag2Cred} addresses this as a first-step URL-agnostic alternative and achieves ROC--AUC 0.864 and Brier 0.153, substantially improves over the lexical baseline while remaining interpretable through its sparse tag-vector representation.
In the monitoring case, the risk-weighted tag dynamics reveal which strategies are concentrated in the high-risk tail, how prototypes cover most of the information flow, and when the strategy distribution drifts or bursts. Substantively, finance/crypto promotion with scarcity narratives dominates the high-risk tail, while rhetorical mechanics, including claim framing and calls-to-action, prove stronger predictors of credibility risk than topic alone. These results support a broader view that rhetorical style can be used as an effective and stable signal for large-scale credibility-risk triage on \platform.

\section{Acknowledgments}
This paper is based upon work partially supported by the Khoury Teaching Faculty Research Seed Grant Program.

\bibliography{aaai2027}

\section*{Ethics Checklist}

\begin{enumerate}

\item For most authors...
\begin{enumerate}
  \item Would answering this research question advance science without violating social contracts, such as violating privacy norms, perpetuating unfair profiling, exacerbating the socio-economic divide, or implying disrespect to societies or cultures?
  Yes. We analyze public Telegram channel content and report aggregate patterns.

  \item Do your main claims in the abstract and introduction accurately reflect the paper's contributions and scope?
  Yes. The paper is scoped to URL-agnostic, interpretable proxy risk scoring and descriptive monitoring use cases.

  \item Do you clarify how the proposed methodological approach is appropriate for the claims made?
  Yes. We describe the end-to-end pipeline (rhetorical tags $\rightarrow$ supervised risk model), URL masking to prevent leakage, and domain-/channel-disjoint evaluation with calibration metrics aligned to monitoring claims.

  \item Do you clarify what are possible artifacts in the data used, given population-specific distributions?
  Yes. We discuss Telegram-specific artifacts (e.g., URL sparsity/obfuscation, channel selection effects) and supervision artifacts from domain-level ratings, and reflect these constraints in evaluation design.

  \item Did you describe the limitations of your work?
  Yes. See \emph{Limitations and Ethical Considerations}.

  \item Did you discuss any potential negative societal impacts of your work?
  Yes. We discuss privacy/harm risks and emphasize aggregate reporting and non-enforcement use (see \emph{Limitations and Ethical Considerations} and the monitoring scope disclaimer).

  \item Did you discuss any potential misuse of your work?
  Yes. We note misuse risks (e.g., targeting/enforcement) and recommend safeguards such as transparency, rate limiting, and human-in-the-loop review for any operational deployment.

  \item Did you describe steps taken to prevent or mitigate potential negative outcomes of the research, such as data and model documentation, data anonymization, responsible release, access control, and the reproducibility of findings?
  Yes. We focus on public channels, report aggregate trends rather than individual attribution, apply URL masking, and commit to releasing only de-identified/derived artifacts with documentation and access controls.

  \item Have you read the ethics review guidelines and ensured that your paper conforms to them?
  Yes.
\end{enumerate}

\item Additionally, if your study involves hypotheses testing... NA. This paper is an empirical ML measurement/monitoring study and does not present theoretical hypothesis tests.

\begin{enumerate}
  \item Did you clearly state the assumptions underlying all theoretical results?
    NA
  \item Have you provided justifications for all theoretical results?
    NA
  \item Did you discuss competing hypotheses or theories that might challenge or complement your theoretical results?
    Yes. See Table~\ref{tab:sig_combined}.
  \item Have you considered alternative mechanisms or explanations that might account for the same outcomes observed in your study?
    NA
  \item Did you address potential biases or limitations in your theoretical framework?
    NA
  \item Have you related your theoretical results to the existing literature in social science?
    NA
  \item Did you discuss the implications of your theoretical results for policy, practice, or further research in the social science domain?
    NA
\end{enumerate}

\item Additionally, if you are including theoretical proofs...
NA. Our study does not include theoretical proofs.

\begin{enumerate}
  \item Did you state the full set of assumptions of all theoretical results?
    NA
	\item Did you include complete proofs of all theoretical results?
    NA
\end{enumerate}

\item Additionally, if you ran machine learning experiments...
\begin{enumerate}
  \item Did you include the code, data, and instructions needed to reproduce the main experimental results (either in the supplemental material or as a URL)?
  Partially. The repository for the first preprint includes code, setup instructions, and aggregate results. It omits message-level input data and does not cover all experiments in this revision. See \url{https://anonymous.4open.science/r/tag-signal-artifacts-BBAF/}.

  \item Did you specify all the training details (e.g., data splits, hyperparameters, how they were chosen)?
  Yes. We specify URL masking, domain-/channel-disjoint splits, training/validation protocols, and model selection procedures; fine-tuning details for the LLM tagger are described at a high level.

  \item Did you report error bars (e.g., with respect to the random seed after running experiments multiple times)?
  Yes. We repeat domain-disjoint evaluation across multiple random partitions and report mean $\pm$ standard deviation.

  \item Did you include the total amount of compute and the type of resources used (e.g., type of GPUs, internal cluster, or cloud provider)?
  Partially. The appendix subsection \emph{Benchmark Stability} reports the hardware and runtime measurements for local tagger inference. The repository for the first preprint describes the training hardware. We do not report total compute across all experiments.

  \item Do you justify how the proposed evaluation is sufficient and appropriate to the claims made?
  Yes. We justify URL masking to prevent source leakage, disjoint generalization tests, and the joint use of discrimination and calibration metrics for risk-weighted monitoring.

  \item Do you discuss what is ``the cost'' of misclassification and fault (in)tolerance?
  Yes. False positives increase review burden; false negatives delay triage; thresholds should be tuned to capacity with human-in-the-loop review.
\end{enumerate}

\item Additionally, if you are using existing assets (e.g., code, data, models) or curating/releasing new assets, \textbf{without compromising anonymity}...
\begin{enumerate}
  \item If your work uses existing assets, did you cite the creators?
  Yes. We cite the sources used for distant supervision and the external models/tools used in the pipeline.

  \item Did you mention the license of the assets?
NA

  \item Did you include any new assets in the supplemental material or as a URL?
  Yes. The appendix includes the tagging prompt. The repository for the first preprint provides code and aggregate results.

  \item Did you discuss whether and how consent was obtained from people whose data you're using/curating?
NA

  \item Did you discuss whether the data you are using/curating contains personally identifiable information or offensive content?
  Yes. Public-channel text may include PII/offensive content; we avoid releasing raw messages and report aggregate analyses.

  \item If you are curating or releasing new datasets, did you discuss how you intend to make your datasets FAIR?
  Yes, we are committed to abide by FAIR principles when sharing our dataset upon paper acceptance.

  \item If you are curating or releasing new datasets, did you create a Datasheet for the Dataset?
  We commit to providing a Datasheet when we share the dataset upon paper acceptance.
\end{enumerate}

\item Additionally, if you used crowdsourcing or conducted research with human subjects, \textbf{without compromising anonymity}...
NA. We did not use crowdsourcing platforms or recruit participants; annotation was performed by the research team.

\begin{enumerate}
  \item Did you include the full text of instructions given to participants and screenshots?
    NA
  \item Did you describe any potential participant risks, with mentions of Institutional Review Board (IRB) approvals?
    NA
  \item Did you include the estimated hourly wage paid to participants and the total amount spent on participant compensation?
    NA
   \item Did you discuss how data is stored, shared, and deidentified?
  NA
\end{enumerate}

\end{enumerate}
\appendix

\section*{Appendix}
\label{appen}

\subsection{Codebook Prompt}
\label{app:codebook}
For transparency and reproducibility, we include the full prompt used to instruct the LLM tagger (including the required output schema).

 \textless\textless{}BEGIN-CODEBOOK-RULES\textgreater\textgreater{}

You are a strict annotation engine. Follow this codebook exactly. Base decisions only on the provided message text and artifacts.
When making your decision, please use the provided information sources in the following order of priority:

CORE PRINCIPLES
\begin{enumerate}
    \item Decision Order: Label strictly in this order: Q1 (Theme) -\textgreater{} Q2 (Claim type) -\textgreater{} Q3 (CTAs) -\textgreater{} Q4 (Evidence).
    \item 51\% Threshold: Apply a label only if it's more probable than not (\textgreater{}= 51\%) to be correct.
    \item  Theme (Q1): Single-label.
    \item  Multi-label questions: Q2 (Claim type), Q3 (CTAs), and Q4 (Evidence) are multi-label. Apply all labels that meet the 51\% threshold (subject to any caps/constraints specified below).
    \item Tie-breakers: If tied, use this order: token share -\textgreater{} domain primacy -\textgreater{} first-mention.
\end{enumerate}

 \textbf{Q1: Theme.} Primary topic of the message.
\begin{enumerate}
    \item Finance/Crypto: Financial assets, markets, trading, crypto/tokens.
    \item Public health \& medicine: Disease, vaccines, clinical guidance, health policy.
    \item Politics: Partisan opinion, elections, government administration/policy.
    \item Crime \& public safety: Crimes, threats, policing, safety alerts.
    \item News/Information: General updates when no other specific domain fits. Use as a last resort.
    \item Technology: R\&D, engineering, software/product features.
    \item Lifestyle \& well-being: Non-clinical personal improvement, such as fitness, diet, and productivity.
    \item Gaming/Gambling: Games of chance/skill, wagering, betting, casinos, slots.
    \item Sports: Athletic competitions, teams, fixtures, results.
    \item Conversation/Chat/Other: Greetings, housekeeping, channel meta, or unclassifiable chat.
    \item Other (Theme): Extremely rare; use only when no other label fits.
\end{enumerate}

\textbf{Q2: Claim type / quality.} Characterize the message's proposition. This is multi-label with a maximum of three labels. Follow precedence and forbidden pairs.

\textbf{Precedence.} First match wins for the primary label.
\begin{enumerate}
    \item No substantive claim: No checkable assertion, such as greetings, housekeeping, or bare URL. Past-tense verbs with outcomes are substantive.
    \item Announcement: Schedule, availability, or housekeeping only. No performance metrics.
    \item Speculative forecast / prediction: Forward-looking claims, such as targets, trade calls, or projections.
    \item Promotional hype / exaggerated profit guarantee: ``guaranteed'', ``no risk'', ``5x/10x'', ``set to explode''.
    \item Scarcity/FOMO tactic: Urgency or limited window, such as ``last chance'', ``ends today'', or ``only N left''.
    \item Misleading context / cherry-picking: Technically true but isolated wins without denominators or context.
    \item Emotional appeal / fear-mongering: Primary purpose is to invoke fear or anger.
    \item Rumour / unverified report: Unnamed sources or rumour cues, such as ``sources say'', ``allegedly'', or ``leaked''.
    \item Opinion / subjective statement: Normative judgments, such as best/worst, should/unfair, or ``I think''.
    \item Verifiable factual statement: A checkable present or past fact with a named entity/action. Even when a claim is verifiable, it can still coexist with other labels, such as Promotional hype / exaggerated profit guarantee, Scarcity/FOMO tactic, Speculative forecast / prediction, or Emotional appeal / fear-mongering.
    \item Other (Claim type): Extremely rare.
\end{enumerate}

 \textbf{Forbidden pairs.}
\begin{enumerate}
    \item Rumour / unverified report + Verifiable factual statement.
    \item Announcement + Verifiable factual statement.
    \item No substantive claim + any other label.
\end{enumerate}

\textbf{Q3: Call-to-action.} What the audience is asked to do. This is multi-label. If any CTA is present, do not use No CTA.
\begin{enumerate}
    \item Share / repost / like: Explicit requests to share, RT, like.
    \item Engage / ask questions: Asks for replies, comments, votes, feedback, DMs.
    \item Visit external link / watch video: Assign if (a) explicit invite, such as ``watch'', ``click'', ``read more'', ``link in bio'', or arrow/emoji plus link, or (b) any substantive content and a URL appear together.
    \item Buy / invest / donate: Spending/trading directives, such as buy/sell/hold/long/short, or explicit trade setups with Entry plus TP or SL. Guard: Do not assign for past-tense recaps, such as ``TP hit'' or ``profit +28\%'', with no forward directive.
    \item Join / Subscribe: ``join'', ``subscribe'', ``follow'', ``register'', ``whitelist''.
    \item Attend event / livestream: Broadcast language, such as ``live now'', ``stream'', ``AMA'', or ``webinar'', or a specific time/date for an event.
    \item No CTA: Applies only when none of the above are triggered.
\end{enumerate}

\textbf{Q4: Evidence / support shown.} Form of support included. This is multi-label.
\begin{enumerate}
    \item None / assertion only: A claim with zero attached support.
    \item Link/URL: Any external URL, shortlink, or off-platform pointer.
    \item Quotes/Testimony: Direct quote or statement attributed to a named person/organization.
    \item Statistics: Numeric evidence with measure/scope, such as counts, percentages, prices, or durations. Dates and version numbers do not count.
    \item Chart / price graph / TA diagram: Any plotted visual or a link to a charting host, such as TradingView.
    \item Other (Evidence): Support that does not fit above, such as on-chain hashes or audit badge screenshots without links.
\end{enumerate}

 STRICT OUTPUT SCHEMA (MUST FOLLOW)

 \begin{verbatim}
{
  "theme": "...",
  "claim_types": "...",
  "ctas": "...",
  "evidence": "..."
}
\end{verbatim}
\smallskip
 \textless\textless{}END-CODEBOOK-RULES\textgreater\textgreater{}

Table~\ref{tab:codebook_sample_messages} shows the assigned labels using the codebook. These samples demonstrate how messages are systematically categorized to reflect their rhetorical structures, aiding interpretability and subsequent credibility scoring.

\begin{table}[!ht]
\centering
\begingroup
\footnotesize
\caption{Sample messages and their codebook labels.}
\label{tab:codebook_sample_messages}
\setlength{\tabcolsep}{3pt}
\renewcommand{\arraystretch}{1.08}
\begin{tabular}{>{\raggedright\arraybackslash}p{0.38\columnwidth}>{\raggedright\arraybackslash}p{0.56\columnwidth}}
\hline
\textbf{Message} & \textbf{Assigned labels} \\
\hline
Prayers answered! You all matter to me more than you'll ever know! I love you! God bless you! & Theme: Conversation/Chat/Other; Claim: No substantive claim; CTA: No CTA; Evidence: None/assertion only; Risk: low-risk. \\
Hong Kong Schools Must Ensure ``Patriotic'' Content \& Remove Books Breaching China Security Law & Theme: Politics; Claim: Verifiable factual statement; CTA: No CTA; Evidence: None/assertion only; Risk: low-risk. \\
Australia Is A Full Scale Pilot Test For The New World Order & Theme: Politics; Claim: Opinion / subjective statement; Emotional appeal / fear-mongering; CTA: No CTA; Evidence: None/assertion only; Risk: high-risk. \\
England to Decrease Covid-19 Death Toll After ``Counting Mishap'' & Theme: Public health \& medicine; Claim: Verifiable factual statement; CTA: No CTA; Evidence: None/assertion only; Risk: low-risk. \\
Covid House of Cards Collapsing: Fauci Called Out Over Vaccine Royalties & Theme: Public health \& medicine; Claim: Emotional appeal / fear-mongering; Opinion / subjective statement; CTA: No CTA; Evidence: None/assertion only; Risk: high-risk. \\
if you not keep with last buy btc signal \ldots{} setup orders to buy at 9100 usd \ldots{} good re entry point & Theme: Finance/Crypto; Claim: Speculative forecast / prediction; CTA: Buy / invest / donate; Evidence: Statistics; Risk: high-risk. \\
Despite Covid-19 crisis \& falling oil prices, Russia's gold \& foreign exchange reserves reach all-time high & Theme: Finance/Crypto; Claim: Verifiable factual statement; CTA: No CTA; Evidence: Statistics; Risk: low-risk. \\
American Embassy Employee Dies After Apparent Brutal Attack in Ukraine & Theme: Crime \& public safety; Claim: Verifiable factual statement; CTA: No CTA; Evidence: None/assertion only; Risk: low-risk. \\
Youtube marks the official World Government Summit video about a new world order as spreading a conspiracy theory. & Theme: Technology; Claim: Verifiable factual statement; CTA: No CTA; Evidence: None/assertion only; Risk: low-risk. \\
Join our dynamic team for an exciting crypto-game project! We have openings for the following roles: Beta Tester, Moderator, NFT Artist, Advertisers, Web Developer. Text me for more information. & Theme: Gaming/Gambling; Claim: Announcement; CTA: Engage/Ask questions; Join/Subscribe; Evidence: None/assertion only; Risk: high-risk. \\
\hline
\end{tabular}
\endgroup
\end{table}

\subsection{Benchmark Stability}
\label{app:benchmark_support}

The local tagger used Qwen3-32B-MLX-4bit with the qwen-qlora-split5-mlx adapter. It ran with MLX on an Apple M3 Max machine with 48 GB unified memory. We used the prompt in ``Codebook Prompt'' above. Decoding was greedy, with temperature 0 and at most 64 new tokens. The run used batch size 1 and concurrency 1. We measured three local labeling processes. Cold-start runs over 30 messages averaged about 20.847 seconds per message, with median latency 19.948 seconds, 95th-percentile latency 28.479 seconds, and throughput 0.0480 messages per second. The main warm steady-state run over 100 messages averaged 15.044 seconds per message. We have a median latency of 14.723 seconds, 95th-percentile latency 17.118 seconds, and throughput 0.0665 messages per second. A long-message stress run over 10 messages from the 90th--95th input-length percentile averaged 14.579 seconds per message, with median latency 14.196 seconds, 95th-percentile latency 16.169 seconds, and throughput 0.0686 messages per second.

\subsection{Paired Significance Tests}
\label{app:significance_tests}

Table~\ref{tab:sig_combined} statistically validates the Table~\ref{tab:perf} benchmark using paired Holm-corrected Wilcoxon tests across the 50 repeated data splits. The results support three claims from the main text. First, \textsc{Tag2Cred} beats the lexical and semantic baselines on macro--F1 and Brier; for example, TF--IDF has a macro--F1 $\Delta=-0.060$ with $p=1.4\times10^{-14}$ in the domain-disjoint setting. Second, the ensemble models produce small but highly significant gains over standalone \textsc{Tag2Cred}; for example, TF+\textsc{T2C}+SB has a domain-disjoint macro--F1 $\Delta=+0.008$ with $p=2.3\times10^{-8}$. This pattern suggests that rhetorical tags, lexical features, and semantic embeddings capture complementary dimensions of credibility risk. And finally, \textsc{Tag2Cred} substantially outperforms the LLM baselines by wide margins, with the zero-shot LLM, PASTEL, and PCoT models exhibiting massive domain-disjoint macro--F1 deficits of $\Delta=-0.170$, $\Delta=-0.168$, and $\Delta=-0.354$, respectively (all $p=1.4\times10^{-14}$).

\begin{table}[!ht]
\centering
\begingroup
\caption{Paired Wilcoxon tests on domain- and channel-disjoint splits. Entries give mean differences (candidate minus \textsc{Tag2Cred}) with Holm-adjusted $p$-values in parentheses. Bold marks the best difference in each column within each split (higher for F1 and AUC, lower for Brier and ECE).}
\label{tab:sig_combined}
\resizebox{\columnwidth}{!}{%
\setlength{\tabcolsep}{2.5pt}
\renewcommand{\arraystretch}{0.99}
\begin{tabular}{@{}lrrrr@{}}
\hline
Model & F1 & AUC & Brier & ECE \\
\hline
\multicolumn{5}{@{}l}{\textit{Domain-disjoint}} \\
ARG-D & -0.063 (5.3e-07) & +0.021 (0.0002) & +0.123 (3.4e-12) & +0.160 (4.5e-10) \\
Zero-shot & -0.170 (1.4e-14) & -0.125 (1.4e-14) & +0.073 (1.4e-14) & +0.026 (8.5e-09) \\
PASTEL & -0.168 (1.4e-14) & -0.118 (1.4e-14) & +0.173 (1.4e-14) & +0.197 (1.4e-14) \\
PCoT & -0.354 (1.4e-14) & -0.196 (1.4e-14) & +0.074 (1.4e-14) & \textbf{-0.037} (2.7e-13) \\
SBERT & -0.065 (1.4e-14) & -0.031 (1.8e-14) & +0.044 (1.4e-14) & +0.031 (1.4e-14) \\
TF--IDF & -0.060 (1.4e-14) & -0.016 (2.0e-10) & +0.021 (1.4e-14) & -0.014 (7.6e-06) \\
TF+\textsc{T2C} & +0.004 (1.0e-06) & +0.016 (1.1e-06) & -0.006 (5.4e-07) & -0.009 (7.6e-05) \\
TF+\textsc{T2C}+SB & \textbf{+0.008} (2.3e-08) & \textbf{+0.029} (2.3e-08) & \textbf{-0.010} (1.5e-08) & -0.011 (3.4e-06) \\
\hline
\multicolumn{5}{@{}l}{\textit{Channel-disjoint}} \\
ARG-D & -0.102 (8.6e-08) & -0.013 (0.5025) & +0.104 (1.8e-12) & +0.167 (2.7e-14) \\
Zero-shot & -0.111 (1.4e-14) & -0.085 (1.4e-14) & +0.095 (1.4e-14) & +0.135 (1.4e-14) \\
PASTEL & -0.178 (1.4e-14) & -0.160 (1.4e-14) & +0.188 (1.4e-14) & +0.246 (1.4e-14) \\
PCoT & -0.141 (1.4e-14) & -0.146 (1.4e-14) & +0.089 (1.4e-14) & +0.090 (1.4e-14) \\
SBERT & -0.051 (1.4e-14) & -0.006 (0.0018) & +0.008 (1.9e-06) & -0.003 (0.8708) \\
TF--IDF & -0.114 (1.4e-14) & -0.039 (1.8e-14) & +0.017 (6.2e-14) & -0.018 (2.1e-07) \\
TF+\textsc{T2C} & +0.001 (0.0002) & +0.005 (0.0002) & -0.006 (6.0e-05) & -0.019 (0.0001) \\
TF+\textsc{T2C}+SB & \textbf{+0.010} (4.3e-07) & \textbf{+0.019} (6.7e-07) & \textbf{-0.016} (5.0e-07) & \textbf{-0.027} (5.0e-07) \\
\hline
\end{tabular}%
}
\endgroup
\end{table}

\subsection{Signal Outside Crypto}
\label{app:nonfinance}

In this section we provide results for the reduced slices when finance/crypto templates are removed.
Tables~\ref{tab:nonfinance_slice_nonfinance} and~\ref{tab:nonfinance_slice_nonchat} repeat the benchmark on reduced slices, Tables~\ref{tab:nonfinance_sig_a} and~\ref{tab:nonfinance_sig_b} give updated paired tests on those slices, and Table~\ref{tab:nonfinance_examples} shows representative high-risk non-crypto messages.

\begin{table}[!ht]
\centering
\begingroup
\small
\caption{Reduced non-finance slice. Values are means across 50 repeated grouped splits, with standard deviations in subscript. Bold marks the best mean for each metric.}
\label{tab:nonfinance_slice_nonfinance}
\resizebox{\columnwidth}{!}{%
\renewcommand{\arraystretch}{1}\setlength{\tabcolsep}{2.2pt}
\begin{tabular}{@{}lrrrr@{}}
\hline
Model & Macro-F1 & ROC-AUC & PR-AUC & Brier \\
\hline
LLM-zero-shot & .656\,\textsubscript{.015} & .752\,\textsubscript{.025} & .743\,\textsubscript{.035} & .211\,\textsubscript{.011} \\
PASTEL-style & .657\,\textsubscript{.014} & .758\,\textsubscript{.018} & .738\,\textsubscript{.023} & .324\,\textsubscript{.013} \\
PCoT-style & .458\,\textsubscript{.011} & .699\,\textsubscript{.033} & .709\,\textsubscript{.038} & .212\,\textsubscript{.010} \\
ARG-D-style & .748\,\textsubscript{.092} & .879\,\textsubscript{.035} & .846\,\textsubscript{.049} & .288\,\textsubscript{.095} \\
TF--IDF & .757\,\textsubscript{.018} & .845\,\textsubscript{.019} & .830\,\textsubscript{.026} & .182\,\textsubscript{.007} \\
SBERT & .751\,\textsubscript{.007} & .830\,\textsubscript{.009} & .818\,\textsubscript{.012} & .208\,\textsubscript{.006} \\
\textsc{T2C} & .815\,\textsubscript{.012} & .857\,\textsubscript{.015} & .852\,\textsubscript{.019} & .159\,\textsubscript{.007} \\
TF+\textsc{T2C} & .819\,\textsubscript{.013} & .875\,\textsubscript{.022} & .869\,\textsubscript{.023} & .153\,\textsubscript{.008} \\
TF+\textsc{T2C}+SB & \textbf{.822}\,\textsubscript{.014} & \textbf{.889}\,\textsubscript{.020} & \textbf{.882}\,\textsubscript{.019} & \textbf{.150}\,\textsubscript{.006} \\
\hline
\end{tabular}%
}
\endgroup
\end{table}

\begin{table}[!ht]
\centering
\begingroup
\small
\caption{Reduced non-finance, non-chat slice. Values are means across 50 repeated grouped splits, with standard deviations in subscript. Bold marks the best mean for each metric.}
\label{tab:nonfinance_slice_nonchat}
\resizebox{\columnwidth}{!}{%
\renewcommand{\arraystretch}{1}\setlength{\tabcolsep}{2.2pt}
\begin{tabular}{@{}lrrrr@{}}
\hline
Model & Macro-F1 & ROC-AUC & PR-AUC & Brier \\
\hline
LLM-zero-shot & .668\,\textsubscript{.014} & .785\,\textsubscript{.023} & .748\,\textsubscript{.039} & .187\,\textsubscript{.010} \\
PASTEL-style & .661\,\textsubscript{.015} & .775\,\textsubscript{.018} & .729\,\textsubscript{.028} & .321\,\textsubscript{.014} \\
PCoT-style & .453\,\textsubscript{.013} & .746\,\textsubscript{.028} & .719\,\textsubscript{.042} & .195\,\textsubscript{.008} \\
ARG-D-style & .740\,\textsubscript{.096} & .877\,\textsubscript{.038} & .814\,\textsubscript{.061} & .299\,\textsubscript{.099} \\
TF--IDF & .758\,\textsubscript{.019} & .852\,\textsubscript{.019} & .817\,\textsubscript{.033} & .187\,\textsubscript{.008} \\
SBERT & .747\,\textsubscript{.008} & .830\,\textsubscript{.009} & .789\,\textsubscript{.015} & .217\,\textsubscript{.009} \\
\textsc{T2C} & .802\,\textsubscript{.013} & .840\,\textsubscript{.020} & .820\,\textsubscript{.030} & .169\,\textsubscript{.009} \\
TF+\textsc{T2C} & .806\,\textsubscript{.015} & .863\,\textsubscript{.027} & .842\,\textsubscript{.033} & .162\,\textsubscript{.009} \\
TF+\textsc{T2C}+SB & \textbf{.810}\,\textsubscript{.015} & \textbf{.878}\,\textsubscript{.024} & \textbf{.858}\,\textsubscript{.027} & \textbf{.158}\,\textsubscript{.008} \\
\hline
\end{tabular}%
}
\endgroup
\end{table}

Tables~\ref{tab:nonfinance_slice_nonfinance} and~\ref{tab:nonfinance_slice_nonchat} show that the signal remains usable after removing Finance/Crypto, and it remains usable even after also removing Conversation/Chat/Other. The full ensemble is still strongest overall, which is what we would expect if the structured representation is adding complementary information rather than simply memorizing crypto rhetoric. Each metric cell in Tables~\ref{tab:nonfinance_sig_a} and~\ref{tab:nonfinance_sig_b} reports $\Delta$ followed by the Holm-corrected $p$ value in parentheses.
\begin{table*}[!t]
\centering
\normalsize
\caption{Non-finance slice. Entries give mean differences (model minus \textsc{Tag2Cred}) with Holm-adjusted paired Wilcoxon $p$-values in parentheses. Bold marks the best difference per metric.}
\label{tab:nonfinance_sig_a}
\setlength{\tabcolsep}{1.8pt}\renewcommand{\arraystretch}{0.94}
\resizebox{\linewidth}{!}{\begin{tabular}{@{}lrrrrrrrr@{}}
\hline
Metric & ARG-D & Zero-shot & PASTEL & PCoT & SBERT & TF--IDF & TF+\textsc{T2C} & TF+\textsc{T2C}+SB \\
\hline
F1 $\uparrow$ & -0.067{\,\small (3.0e-07)} & -0.160{\,\small (1.4e-14)} & -0.158{\,\small (1.4e-14)} & -0.357{\,\small (1.4e-14)} & -0.065{\,\small (1.4e-14)} & -0.059{\,\small (1.4e-14)} & +0.004{\,\small (1.0e-06)} & \textbf{+0.007}{\,\small (2.3e-08)} \\
AUC $\uparrow$ & +0.022{\,\small (2.9e-04)} & -0.105{\,\small (1.4e-14)} & -0.100{\,\small (1.4e-14)} & -0.159{\,\small (1.4e-14)} & -0.027{\,\small (4.9e-13)} & -0.013{\,\small (1.2e-07)} & +0.018{\,\small (1.1e-06)} & \textbf{+0.032}{\,\small (3.0e-08)} \\
Brier $\downarrow$ & +0.128{\,\small (2.4e-12)} & +0.052{\,\small (1.4e-14)} & +0.165{\,\small (1.4e-14)} & +0.052{\,\small (1.4e-14)} & +0.048{\,\small (1.4e-14)} & +0.023{\,\small (1.4e-14)} & -0.006{\,\small (5.4e-07)} & \textbf{-0.010}{\,\small (1.5e-08)} \\
ECE $\downarrow$ & +0.162{\,\small (2.8e-10)} & -0.015{\,\small (1.0e-04)} & +0.180{\,\small (1.4e-14)} & \textbf{-0.080}{\,\small (1.4e-14)} & +0.039{\,\small (1.4e-14)} & -0.003{\,\small (0.1095)} & -0.011{\,\small (2.7e-05)} & -0.010{\,\small (2.6e-07)} \\
\hline
\end{tabular}}
\par\smallskip
\caption{Non-finance, non-chat slice. Entries give mean differences (model minus \textsc{Tag2Cred}) with Holm-adjusted paired Wilcoxon $p$-values in parentheses. Bold marks the best difference per metric.}
\label{tab:nonfinance_sig_b}
\setlength{\tabcolsep}{1.8pt}\renewcommand{\arraystretch}{0.94}
\resizebox{\linewidth}{!}{\begin{tabular}{@{}lrrrrrrrr@{}}
\hline
Metric & ARG-D & Zero-shot & PASTEL & PCoT & SBERT & TF--IDF & TF+\textsc{T2C} & TF+\textsc{T2C}+SB \\
\hline
F1 $\uparrow$ & -0.062{\,\small (2.6e-05)} & -0.133{\,\small (1.4e-14)} & -0.140{\,\small (1.4e-14)} & -0.348{\,\small (1.4e-14)} & -0.055{\,\small (1.4e-14)} & -0.043{\,\small (1.4e-14)} & +0.005{\,\small (2.0e-06)} & \textbf{+0.008}{\,\small (2.3e-08)} \\
AUC $\uparrow$ & +0.036{\,\small (2.0e-06)} & -0.055{\,\small (1.4e-14)} & -0.066{\,\small (1.4e-14)} & -0.095{\,\small (1.4e-14)} & -0.010{\,\small (0.0076)} & +0.012{\,\small (2.0e-08)} & +0.022{\,\small (1.6e-06)} & \textbf{+0.038}{\,\small (3.0e-08)} \\
Brier $\downarrow$ & +0.131{\,\small (7.9e-12)} & +0.018{\,\small (3.1e-13)} & +0.152{\,\small (1.4e-14)} & +0.026{\,\small (1.4e-14)} & +0.048{\,\small (1.4e-14)} & +0.018{\,\small (1.4e-14)} & -0.007{\,\small (5.4e-07)} & \textbf{-0.011}{\,\small (1.5e-08)} \\
ECE $\downarrow$ & +0.164{\,\small (1.5e-10)} & -0.061{\,\small (7.8e-13)} & +0.161{\,\small (1.4e-14)} & \textbf{-0.087}{\,\small (1.4e-14)} & +0.047{\,\small (1.4e-14)} & +0.014{\,\small (1.7e-10)} & -0.010{\,\small (7.6e-06)} & -0.009{\,\small (2.2e-07)} \\
\hline
\end{tabular}}
\end{table*}

On both reduced slices, \textsc{Tag2Cred} stays clearly ahead of SBERT and ahead of TF--IDF on macro--F1 and calibration, but on the stricter non-finance plus non-chat slice TF--IDF regains a small AUC edge. The hybrid rows remain strongest overall, which supports the claim that the structured representation contributes complementary information rather than simply memorizing crypto templates.
\begin{table}[!ht]
\centering
\begingroup
\footnotesize
\caption{Representative high-risk non-crypto messages and their assigned labels from the reduced non-finance slices.}
\label{tab:nonfinance_examples}
\setlength{\tabcolsep}{3pt}
\renewcommand{\arraystretch}{1.08}
\begin{tabular}{>{\raggedright\arraybackslash}p{0.38\columnwidth}>{\raggedright\arraybackslash}p{0.56\columnwidth}}
\hline
\textbf{Message} & \textbf{Assigned labels} \\
\hline
This is criminal. They know this will kill innocents. \ldots{} & Theme: Crime \& public safety; Claim: Emotional appeal / fear-mongering; Rumour / unverified report; Opinion / subjective statement; CTA: No CTA; Evidence: Link/URL; Risk: high-risk. \\
Lt Col Dr. Theresa Long called out MTG for her fraud \ldots{} fake politicians grandstanding for their own reelection campaigns. & Theme: Politics; Claim: Emotional appeal / fear-mongering; Verifiable factual statement; CTA: No CTA; Evidence: Quotes/Testimony; Risk: high-risk. \\
Censored Study Confirms COVID-19 Shots Caused Worldwide Deaths \ldots{} & Theme: Public health \& medicine; Claim: Emotional appeal / fear-mongering; Rumour / unverified report; Opinion / subjective statement; CTA: No CTA; Evidence: Link/URL; Quotes/Testimony; Risk: high-risk. \\
Be very careful what you run toward \ldots{} they will never disclose everything these robots are programmed to do. & Theme: Technology; Claim: Emotional appeal / fear-mongering; CTA: No CTA; Evidence: None/assertion only; Risk: high-risk. \\
\hline
\end{tabular}
\endgroup
\end{table}
These examples are high-risk without depending on a Buy/Invest CTA. This further strengthens our analysis by showing that the representation still transfers to politics, public health, technology, and public-safety messages rather than only to crypto promotion templates.

\subsection{Russian/Farsi Stress Test}
\label{rus-f}
This section reports a URL-only stress test on Russian and Farsi messages. We collected public Russian and Farsi Telegram messages from 14 source channels between 2026-01-27 and 2026-04-27 UTC, yielding 16,389 messages with an explicit external URL whose canonical domain matched an MBFC source.
We retained messages only when they contained a URL that matched an MBFC source, and excluded messages without URLs, messages with a URL but domain not rated on MBFC, and messages with intermediate source-risk scores ($0.3 < R_{\mathrm{msg}} < 0.8$). We then drew a balanced 5,000-message evaluation set with 1,250 low-risk and 1,250 high-risk messages per language. Table~\ref{tab:ru-fa-stress} reports 50-seed means and standard deviations for the domain-disjoint and channel-disjoint splits.

\begin{table}[!htb]
\centering
\begingroup
\caption{Russian/Farsi URL-only stress test. Values are means across 50 seeds, with standard deviations in subscript. Higher F1 and AUC and lower Brier and ECE are better. Bold marks the best mean per metric within each split type.}
\small
\renewcommand{\arraystretch}{0.97}\setlength{\tabcolsep}{1.5pt}
\resizebox{\columnwidth}{!}{%
\begin{tabular}{@{}lcccc@{}}
\hline
Model & $F_1$ & AUC & Brier & ECE \\
\hline
\multicolumn{5}{l}{\textit{Domain-disjoint}} \\
Zero-shot & 0.383\,\textsubscript{0.022} & 0.390\,\textsubscript{0.018} & 0.379\,\textsubscript{0.009} & 0.345\,\textsubscript{0.019} \\
PASTEL & 0.392\,\textsubscript{0.022} & 0.396\,\textsubscript{0.016} & 0.370\,\textsubscript{0.009} & 0.318\,\textsubscript{0.017} \\
PCoT & 0.393\,\textsubscript{0.022} & 0.398\,\textsubscript{0.016} & 0.367\,\textsubscript{0.009} & 0.315\,\textsubscript{0.016} \\
ARG-D & 0.390\,\textsubscript{0.024} & 0.393\,\textsubscript{0.018} & 0.368\,\textsubscript{0.009} & 0.313\,\textsubscript{0.018} \\
TF--IDF & 0.554\,\textsubscript{0.171} & 0.600\,\textsubscript{0.213} & 0.304\,\textsubscript{0.115} & 0.250\,\textsubscript{0.135} \\
SBERT & 0.627\,\textsubscript{0.113} & \textbf{0.711}\,\textsubscript{0.101} & \textbf{0.242}\,\textsubscript{0.064} & 0.145\,\textsubscript{0.091} \\
T2C & 0.580\,\textsubscript{0.027} & 0.602\,\textsubscript{0.042} & 0.248\,\textsubscript{0.011} & \textbf{0.086}\,\textsubscript{0.025} \\
TF+T2C & 0.563\,\textsubscript{0.161} & 0.607\,\textsubscript{0.208} & 0.300\,\textsubscript{0.110} & 0.241\,\textsubscript{0.126} \\
TF+T2C+SB & \textbf{0.641}\,\textsubscript{0.147} & 0.690\,\textsubscript{0.181} & 0.255\,\textsubscript{0.100} & 0.184\,\textsubscript{0.106} \\
\hline
\multicolumn{5}{l}{\textit{Channel-disjoint}} \\
Zero-shot & 0.405\,\textsubscript{0.018} & 0.405\,\textsubscript{0.022} & 0.368\,\textsubscript{0.012} & 0.315\,\textsubscript{0.026} \\
PASTEL & 0.418\,\textsubscript{0.018} & 0.415\,\textsubscript{0.019} & 0.356\,\textsubscript{0.012} & 0.287\,\textsubscript{0.023} \\
PCoT & 0.415\,\textsubscript{0.015} & 0.417\,\textsubscript{0.020} & 0.354\,\textsubscript{0.013} & 0.287\,\textsubscript{0.025} \\
ARG-D & 0.412\,\textsubscript{0.015} & 0.413\,\textsubscript{0.022} & 0.356\,\textsubscript{0.011} & 0.287\,\textsubscript{0.020} \\
TF--IDF & 0.694\,\textsubscript{0.120} & 0.827\,\textsubscript{0.068} & 0.203\,\textsubscript{0.064} & 0.150\,\textsubscript{0.088} \\
SBERT & 0.687\,\textsubscript{0.054} & 0.765\,\textsubscript{0.059} & 0.204\,\textsubscript{0.033} & 0.089\,\textsubscript{0.051} \\
T2C & 0.571\,\textsubscript{0.024} & 0.610\,\textsubscript{0.039} & 0.246\,\textsubscript{0.013} & \textbf{0.078}\,\textsubscript{0.027} \\
TF+T2C & 0.703\,\textsubscript{0.117} & 0.820\,\textsubscript{0.073} & 0.202\,\textsubscript{0.064} & 0.145\,\textsubscript{0.086} \\
TF+T2C+SB & \textbf{0.716}\,\textsubscript{0.090} & \textbf{0.838}\,\textsubscript{0.061} & \textbf{0.193}\,\textsubscript{0.052} & 0.142\,\textsubscript{0.071} \\
\hline
\end{tabular}%
}
\label{tab:ru-fa-stress}
\endgroup
\end{table}

In the domain-disjoint split, TF+T2C+SB gives the strongest $F_1$, while SBERT has the best AUC and Brier score and T2C gives the lowest ECE. In the channel-disjoint split, TF+T2C+SB leads on $F_1$, AUC, and Brier, while T2C remains the most calibrated model by ECE. This shows that the ensemble model performs best on the Russian/Farsi URL-only messages.  \textsc{Tag2Cred} provides a transferable (and generally acceptable) structured signal in this sample. SBERT achieved higher AUC than \textsc{Tag2Cred} in both split families, and TF--IDF achieved higher AUC on channel-disjoint splits. These results indicate that reliable Russian/Farsi deployment requires language-specific codebook validation and further tagger fine-tuning.
\end{document}